\documentclass{article}

    \usepackage[final, nonatbib]{neurips_2025}

\usepackage[utf8]{inputenc} 
\usepackage[T1]{fontenc}    
\usepackage{hyperref}       
\usepackage{url}            
\usepackage{booktabs}       
\usepackage{amsfonts}       
\usepackage{nicefrac}       
\usepackage{microtype}      
\usepackage{xcolor}         
\usepackage{graphicx}
\usepackage{titlesec}
\titlespacing{\paragraph}{0pt}{0.0\baselineskip}{1em}
\usepackage{amsmath}

\usepackage{caption}
\usepackage[most]{tcolorbox}
\usepackage{enumitem}
\usepackage{lmodern} 
\tcbset{
  promptbox/.style={
    breakable,
    enhanced,
    colback=black!2,
    colframe=black!35,
    boxrule=0.4pt,
    left=6pt,right=6pt,top=6pt,bottom=6pt,
    sharp corners,
    before skip=8pt, after skip=12pt,
  }
}
\newtcolorbox[auto counter, number within=section]{prompt}[1][]{%
  promptbox,
  fonttitle=\bfseries\ttfamily,
  title={Prompt~\thetcbcounter},
  #1
}

\title{BikeBench: A Bicycle Design Benchmark for Generative Models with Objectives and Constraints}
\makeatletter
\providecommand{\@trackname}{}
\makeatother
\author{%
  Lyle Regenwetter \\
  MIT\\
  \texttt{regenwet@mit.edu} \\
  \And
  Yazan Abu Obaideh\\
  ProgressSoft \\
  \texttt{yazan.amer@protonmail.com} \\
  \And
  Fabien Chiotti \\
  Utrecht University \\
  \texttt{fabien.chiotti@gmail.com} \\
  \And
  Ioanna Lykourentzou \\
  Utrecht University \\
  \texttt{i.lykourentzou@uu.nl} \\
  \And
  Faez Ahmed \\
  MIT \\
  \texttt{faez@mit.edu} \\
}

\begin{document}

\maketitle

\begin{abstract}
    \vspace*{-1mm}
    We introduce BikeBench, an engineering design benchmark for evaluating generative models on problems with multiple real-world objectives and constraints. As generative AI's reach continues to grow, evaluating its capability to understand physical laws, human guidelines, and hard constraints grows increasingly important. Engineering product design lies at the intersection of these difficult tasks, providing new challenges for AI capabilities. BikeBench evaluates AI models' capabilities to generate bicycle designs that not only resemble the dataset, but meet specific performance objectives and constraints. To do so, BikeBench quantifies a variety of human-centered and multiphysics performance characteristics, such as aerodynamics, ergonomics, structural mechanics, human-rated usability, and similarity to subjective text or image prompts. Supporting the benchmark are several datasets of simulation results, a dataset of 10,000 human-rated bicycle assessments, and a synthetically generated dataset of 1.6M designs, each with a parametric, CAD/XML, SVG, and PNG representation. BikeBench is uniquely configured to evaluate tabular generative models, large language models (LLMs), design optimization, and hybrid algorithms side-by-side. Our experiments indicate that LLMs and tabular generative models fall short of hybrid GenAI+optimization algorithms in design quality, constraint satisfaction, and similarity scores, suggesting significant room for improvement. We hope that BikeBench, a first-of-its-kind benchmark, will help catalyze progress in generative AI for constrained multi-objective engineering design problems. We provide code, data, an interactive leaderboard, and other resources at \href{https://github.com/Lyleregenwetter/BikeBench}{https://github.com/Lyleregenwetter/BikeBench}.
    \vspace*{-2mm}
\end{abstract}
\section{Introduction}

\begin{figure}[!htb]
    \centering
    \includegraphics[width=0.85\linewidth]{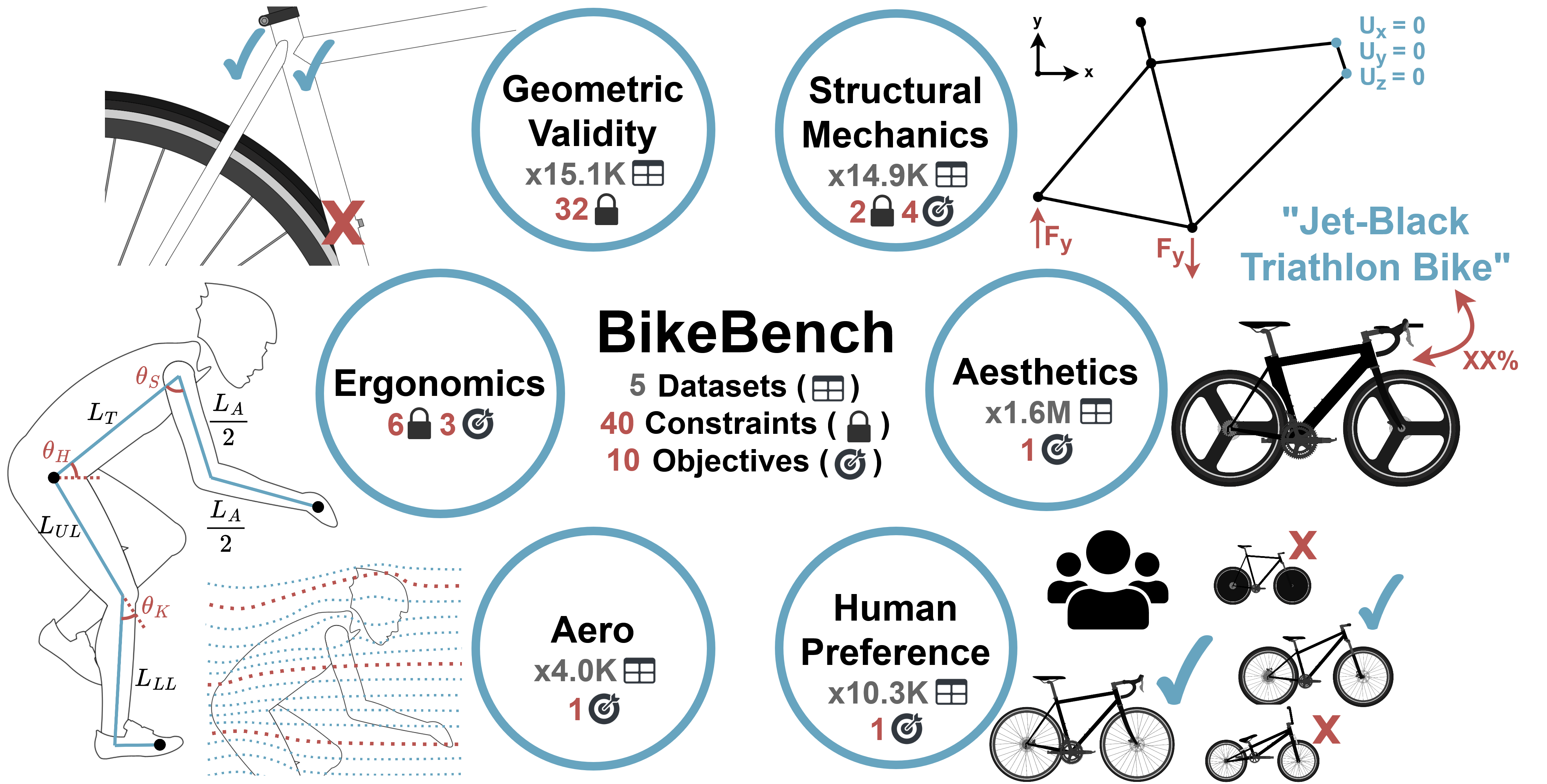}
    \caption{BikeBench is a set of evaluators, metrics, and datasets---built to benchmark generative models' capability to synthesize parametric bicycle designs satisfying a variety of objectives and constraints.}
    \label{fig:overview}
\end{figure}

Generative AI has recently captured widespread attention for its general-purpose problem-solving capabilities~\cite{achiam2023gpt, guo2025deepseek, rombach2022high}. 
Despite a wealth of exploratory work~\cite{oh2019deep, wang2020deep, wu2021deepcad, nobari2024link}, generative AI has not seen widespread adoption~\cite{regenwetter2022deep} in the trillion-dollar engineering design industry~\cite{ACEC2024, RAEng2022}. Engineering design can be generally characterized as the methodical decision-making process needed for the physical realization of products, systems, or other real-world solutions~\cite{beitz1996engineering, buede2024engineering, shigley1985mechanical}. Many of engineering design's requisite skills pose significant challenges for current generative AI models~\cite{faruqi2024shaping, gaier2024generative}, including exact constraint satisfaction, adherence to quantitative and qualitative design guidelines, and intimate knowledge of multidisciplinary physical laws. In ship hull design, for example, generative models have been found to extensively violate geometric, performance, and safety constraints, sometimes over $95\%$ of the time~\cite{regenwetter2024constraining}. In large language model (LLM) benchmarks, models fail to extract precise design regulations from engineering standards~\cite{doris2025designqa}. Finally, in free-form structural design, generative models fall far short of optimization algorithms~\cite{woldseth2022use} due to their inability to learn generalizable physics rules. To maximally leverage generative AI for engineering design, practitioners need models that can precisely achieve constraints, understand both objective and subjective design guidelines, and adhere to physical laws. To help the community assess progression, we introduce a benchmark (Figure~\ref{fig:overview}) focused on bicycle design, a problem that prominently features each of these challenges. 

\vspace{-2mm}

As a constrained, multiphysics-guided, and human-centered design benchmark, BikeBench fills several important voids in the existing space of generative modeling benchmarks. Unlike many image, sketch, or 3D model-based benchmarks~\cite{Ha2017quickdraw, chang2015shapenet} used for design, BikeBench evaluates exclusively parametric designs. This forces benchmarked models to synthesize designs with an exact mapping to a Computer-Aided-Design (CAD) file, essentially guaranteeing a precise, ready-to-manufacture bicycle model, rather than a more abstract image, sketch, or point cloud with less-clear downstream value. BikeBench also differs from CAD datasets, which rarely have features to evaluate multiphysics objectives, hard constraints, or human-centered preferences~\cite{kim2020large, koch2019abc, willis2021fusion}. In contrast, BikeBench is composed of 10 multidisciplinary design objectives and 40 design constraints, all of which revolve around a rich set of design evaluators. These evaluators leverage datasets of physics simulations, a geometry engine and renderer, and even a dataset of real human-sourced design assessments. 

Unlike classic generative modeling benchmarks, BikeBench is more than just an exercise in the maximization of distributional similarity. To measure the practical design capabilities of a generative model, BikeBench also calculates design quality, constraint violation, and diversity scores over sample sets---metrics that are frequently overlooked, despite their outsized importance in engineering design~\cite{regenwetter2023beyond}. For a model to simultaneously succeed in all of BikeBench's metrics, it must strategically deviate from the data manifold to improve performance and constraint satisfaction in targeted ways. BikeBench supports tabular models, LLMs, direct design optimization algorithms, and hybrid algorithms, providing a unique opportunity to benchmark previously disparate algorithms side-by-side. Our benchmarking results suggest that hybrid GenAI+optimization algorithms are current frontrunners, while LLMs have extensive room for growth and improvement in constrained engineering design problems. Key contributions of this paper are summarized as follows:
\begin{itemize}
    \item We introduce a synthetic dataset of 1.6M bicycle designs represented as tabular data, SVG, PNG, and XML files for Computer-Aided-Design (CAD) software. This dataset supports a variety of design generation tasks such as text-to-CAD, image-to-CAD, and parametric-to-image generation. 
    \item We introduce a dataset of 10,000 human-sourced ratings of bicycle designs. These ratings model subjective human assessments of designs' usability. 
    \item We introduce BikeBench, a benchmark evaluating design quality, constraint satisfaction, similarity, and diversity of generative-model-synthesized bicycle designs. The benchmark features 50 design evaluations spanning aerodynamics, structural mechanics, ergonomics, aesthetics, geometric feasibility, and human perception of usability.  
    \item We benchmark a variety of design generation techniques, including GPT-5-high, multiple tabular generative models, optimization-augmented generative models, and both gradient-based and heuristic design optimization algorithms. 
\end{itemize}

\section{Background}
We provide a brief overview of engineering design benchmarks for generative models, as well as benchmarks constructed to evaluate multiple classes of algorithms. A background on data-driven bicycle design is included in Appendix~\ref{appx:bikedesignbackground}, as well as a motivation for the selection of bicycle design as a benchmark problem. 

\paragraph{Engineering Design Benchmarks for Generative Models: }
BikeBench joins a limited set of engineering design benchmarks for the evaluation and comparison of generative models. Some benchmarks evaluate question-answering capability for engineering standards~\cite{doris2025designqa},  simulations~\cite{ezemba2025simulation}, or larger taxonomies of engineering knowledge~\cite{zhou2025engibench}, but these fall short of benchmarking design synthesis capabilities. Engineering design is highly data-scarce~\cite{picard2023dated}, and most datasets are not configured as design synthesis benchmarks~\cite{jmdspecialissue2019}. Regardless, several engineering design datasets can be configured to support performance-aware generative model training~\cite{wollstadt2022carhoods10k, yoo2025deepwheel, whalen2021simjeb, hong2025deepjeb}. While a few of these datasets feature numerous objectives and constraints~\cite{bagazinski2023ship, cobb2023aircraftverse}, most lack standardized evaluation criteria. Dedicated benchmarks for GenAI-based design synthesis have also been introduced, but primarily support just parametric GenAI models~\cite{felten2025engibench}. BikeBench is the first design synthesis benchmark with numerous multidisciplinary objectives and constraints to compare many classes of algorithms against various types of GenAI. 

\paragraph{Benchmarks for multiple classes of algorithms:}
Few benchmarks directly compare generative models to entirely different classes of algorithms, because such comparisons may superficially seem `unfair.' For example, even though generative models more and more frequently compete with optimization algorithms~\cite{shin2023topology}, the comparison may not seem fair because optimization algorithms are allowed to call evaluators, while purely data-driven models must infer from datasets. We contend that such comparisons are rigorous, important, and particularly timely. In practice, generative AI models are regularly placed in direct competition with other classes of algorithms. Direct competition across classes of models can be easily appreciated in the historical development of deep-learning-based computer vision over algorithmic methods~\cite{o2020deep} or LLMs over classical language models~\cite{naseem2021comprehensive, young2018recent}. Benchmarks that embrace this direct competition will help practitioners understand shortcomings and capacity for growth. Such benchmarking is gaining traction~\cite{trabucco2022design}, particularly in fields like engineering design~\cite{masood2024generative}, where the adoption of generative AI has faced considerable resistance due to the strength of alternative methods~\cite{woldseth2022use}. BikeBench helps fill this need for engineering design benchmarks that accommodate different classes of algorithms.

\section{Datasets}
BikeBench introduces several new datasets, many of which are used to train the predictive models that BikeBench uses to evaluate many facets of bicycle design. BikeBench also consolidates and adapts two existing datasets. The \textbf{BIKED dataset}~\cite{regenwetter2022biked}, comprised of 4,500 human-design bikes, serves as the basis for BikeBench's distribution-modeling component. BikeBench uses a custom 64-parameter subset of salient design features from BIKED's full representation, which are described in Appendix~\ref{appx:parlist}. It also uses a more selective curation and deduplication process, resulting in 3,795 designs. The \textbf{FRAMED dataset}~\cite{regenwetter2023framed}, focusing on structural mechanics of bike frames, features nearly 15,000 designs simulated using Finite Element Analysis (FEA) under multiple loading cases. FRAMED supports BikeBench's structural mechanics evaluators and frame validity analysis. We refer readers to the respective papers for more details on BIKED and FRAMED.

\subsection{Dataset: 1.6M Synthetically Generated Bicycle Designs}

We introduce a new dataset of 1.6 million synthetically generated bicycle designs. For each design, we provide parametric data, images, XML files, and, for convenience, CLIP embeddings of the images. We generate this dataset using a \textbf{customized renderer} running parallelized bare-bones instances of the BikeCAD software. 

\begin{figure}[!htb]
    \centering
    \includegraphics[width=\linewidth]{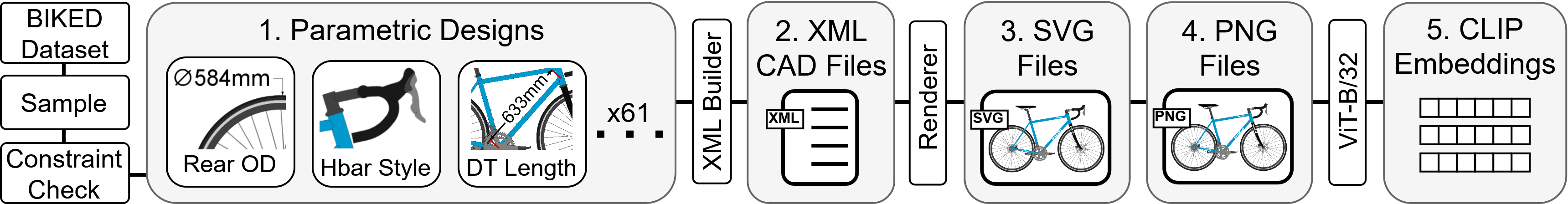}
    \caption{Overview of the synthetic data generation pipeline.}
    \label{fig:data_pipeline}
\end{figure}

Half of the synthetic datapoints are generated to be `realistic' using CTGAN~\cite{xu2019modeling}, a tabular GAN model specializing in realistic synthetic data generation. The other half are generated to be `random' using uniform random sampling from the dataset's most extreme bounds. Each design is procedurally mapped to a Computer-Aided-Design (CAD) file. Though CAD file structure varies widely from software to software, BikeCAD files adopt an XML format and typically consist of several thousand key-value pairs. Our custom renderer is then invoked to render the CAD-XML files into Scalable Vector Graphic (SVG) files, which are then converted to rasterized Portable Network Graphics (PNG) images. Finally, we calculate the CLIP embedding for each generated image using OpenAI's `clip-vit-base-patch32', a ViT-B/32 model provided by HuggingFace. The intermediates (XML, SVG, PNG) are retained and included in the dataset. BikeBench primarily uses this dataset to train a direct embedding model from BikeBench's parametric design space to the CLIP embedding space, allowing for aesthetics-based design evaluation. However, the dataset has broad utility outside of the benchmarking context. It can support a wide variety of predictive modeling and design generation tasks, such as text-to-CAD, image-to-CAD, parametric-to-image, etc. Using the parallelized pipeline, the full 1.6M designs took approximately four days to generate on an ordinary workstation and occupy approximately 750GB of storage, compressed. Users can easily run our pipeline to generate more data at their discretion. 

\subsection{Dataset: 10,000 Human-Sourced Bicycle Ratings}
We also introduce a dataset of 10,350 human ratings of perceived bicycle usability. These ratings were collected through a rigorous selection procedure. Rather than spreading ratings evenly across the 4500 designs in BIKED, we instead opted to focus on a representative subset of 200 designs and collect approximately 50 ratings per bike, allowing us to more easily perform statistical significance tests and calculate rater agreement. This selection process for the 200 designs aimed to achieve a uniform distribution across BikeBench's 64 key parameters. To target users with cycling experience and familiarity with bike usability, only countries with at least a 35\% weekly riding frequency were included, limiting eligibility to 14 countries based on a survey across 28 countries in 2022~\cite{ipsos2022cycling}. Prolific was used to crowdsource participants. These participants answered ``Yes'' or ``No'' to the question: ``Does this bicycle look easy to use?''. Binary (yes/no) assessments were selected to avoid the flaws of continuous ratings~\cite{10.3389/fict.2015.00013}. The 200 sampled bikes were divided into four groups of 50, ensuring diversity within each group. This reduced rater fatigue and helped maintain consistent rating quality throughout the session. After the completion of the rating process, a minimum rating time threshold of 90 seconds was set to remove participants who swiped too quickly, giving little time to fully evaluate each bike design. This reduced the number of valid respondents to an average of 50.75 per group, resulting in 10,350 valid ratings. BikeBench configures this dataset as a regression problem (predicting the proportion of raters who would consider a bike `easy to use'). However, a classification dataset of consensus assessments was also extracted through statistical testing, as described in Appendix~\ref{appx:usabilityclassification}. Information on data collection from human subjects is included in Appendix~\ref{appx:humansubject}. 

\subsection{Dataset: 4K Cyclist Aerodynamics Simulations}
We finally introduce a new dataset composed of 4,000 simulations of 3D cyclist models in various poses. The dataset reports steady-state drag force under a $10,\mathrm{m/s}$ relative headwind, as evaluated by a computational fluid dynamics (CFD) analysis. The 3D cyclist models are parameterized by six anthropometric measurements sampled from an approximate model of published population statistics \cite{Gordon_1989, McDowell_2009, McDowell_2005, Fryar_2016, Fryar_2012}, and five parameters defining the interface points between the cyclist and the bike, which are sampled from BIKED.

\section{Evaluation Criteria}
In this section, we discuss BikeBench's design evaluation criteria. Design problems almost always feature competing objectives. Befittingly, BikeBench's many objectives and constraints are challenging to optimize simultaneously. For example, the most trivial ways to optimize performance objectives will typically involve violating constraints. In this section, we discuss how BikeBench's various datasets and tools are combined to evaluate a medley of functional attributes for bicycle design.

\paragraph{Geometric Feasibility:} 

In data-driven parametric design, highly expressive parametrization schemes often have a drawback in `allowing' invalid configurations. BikeBench's 64 design parameters enable considerable expressive flexibility, but can be selected or generated in ways that result in a variety of geometrically infeasible designs---a common pattern among generative models trained on BIKED~\cite{regenwetter2022design}. These designs commonly feature overlapping or disconnected components, parts with negative dimensions, and frames that violate the triangle inequality, for example. This infeasibility must therefore be identified and evaluated through constraint checks. We have compiled a set of 31 closed-form geometric constraint checks (listed in full in Appendix~\ref{appx:objlist}), which is significantly expanded from a set released in BIKED~\cite{regenwetter2022biked}. We supplement these closed-form checks with a data-driven feasibility predictor trained on FRAMED's binary classification data to identify bike frames with more complex geometric issues. 

\paragraph{Structural Soundness:} 
The structural soundness of a bike's frame plays an important role in its comfort, power-efficiency, and safety. Bike frames are generally preferred to be as rigid as possible, which limits dissipation of energy due to the flexure of the frame when pedaling~\cite{wilson2020bicycling}. We calculate several `composite' structural performance indicators---planar, transverse, and eccentric compliance, as well as frame weight---which are all considered objectives to minimize. We also calculate planar and eccentric safety factor constraints, stating that the yield stress of the frame material must exceed the maximum stress incurred during planar or eccentric loads. We use a model trained on FRAMED~\cite{regenwetter2023framed} to predict these structural attributes. Similar predictive models trained on FRAMED have been shown to closely align with real-world experimental simulations~\cite{regenwetter2023framed}.

\paragraph{Aerodynamics:} 
Aerodynamic drag reduction is a principal consideration in competitive cycling, but is generally beneficial in all cycling settings. In general, the drag force directly incurred by the bicycle is much smaller than the drag incurred by the cyclist's body. Thus, drag is primarily minimized through repositioning of the cyclist, which is a function of the rider's anatomical geometry and the positioning of bicycle components that the rider interfaces with (saddle, handlebars, and pedals). Therefore, aerodynamic performance is a factor of both a bicycle and an associated rider. To quantify aerodynamics, we first calculate the interface points between the bike and the cyclist, then call a predictive model trained on the aerodynamics data for a drag force estimate.    

\paragraph{Ergonomics:} 
In addition to playing a significant role in aerodynamics, cyclist geometry also plays an important role in ergonomics. Examining the range of angles experienced by certain key joints during regular cycling activity~\cite{burt2022bike} is a simple indicator for ergonomic fit. We develop a kinematics solver which calculates a rider's maximum knee angle, hip angle, and shoulder angle and compares them to published target values for various types of cycling~\cite{burt2022bike}. This solver yields six validity constraints and three ergonomic objective scores which are calculated as a function of the bicycle design, the cyclist's anthropometric measurements, and the cyclist's use case (road biking, mountain biking, or commuting).

\paragraph{Human Perception of Usability:} 

Consumers frequently use subjective criteria to evaluate products. Increasingly, products are marketed as `user-friendly,' with usability transitioning from a historic perception as a `bonus' feature to a requisite expectation for many modern products~\cite{jordan2014usability}. Additionally, manufacturers emphasize their products as `people-oriented,' particularly for items that involve direct body contact or require manual operation~\cite{Cheng2011NewTO}. Thus, it is important to focus not only on the performance of the bicycle but also on its perceived usability. Accounting for human-centered criteria allows insights gathered from a broad population to help align bicycles design with growing consumer expectations. We evaluate the perceived usability of bikes using a regression model trained on our new dataset of 10,000 human-sourced usability scores.

\paragraph{Aesthetics:} 
Bicycles are regularly used for fashion, lifestyle, and other cultural statements, elevating the bicycle from a mere transportation product to a versatile tool for nuanced individual expression~\cite{hoor2022bicycle}. As such, the bicycle design industry is heavily influenced by aesthetics, individual preferences, and customization. Because individualized design customization requires significant design throughput, customization is an area where data-driven design methods can particularly excel. Thus, generative models would ideally be able to conditionally generate designs that are customized for an individual user's aesthetic preferences, as communicated through a reference image or a text prompt. To enable this sort of conditioning, we train a predictive model to directly estimate CLIP embeddings of parametric bicycle designs using our dataset of 1.6M synthetic bike designs. By calculating CLIP embeddings for parametric designs, we can quickly evaluate their similarity to text prompts or reference images. 

\section{Benchmarking Metrics and Procedure}
\begin{figure}[!htb]
    \centering
    \includegraphics[width=0.9\linewidth]{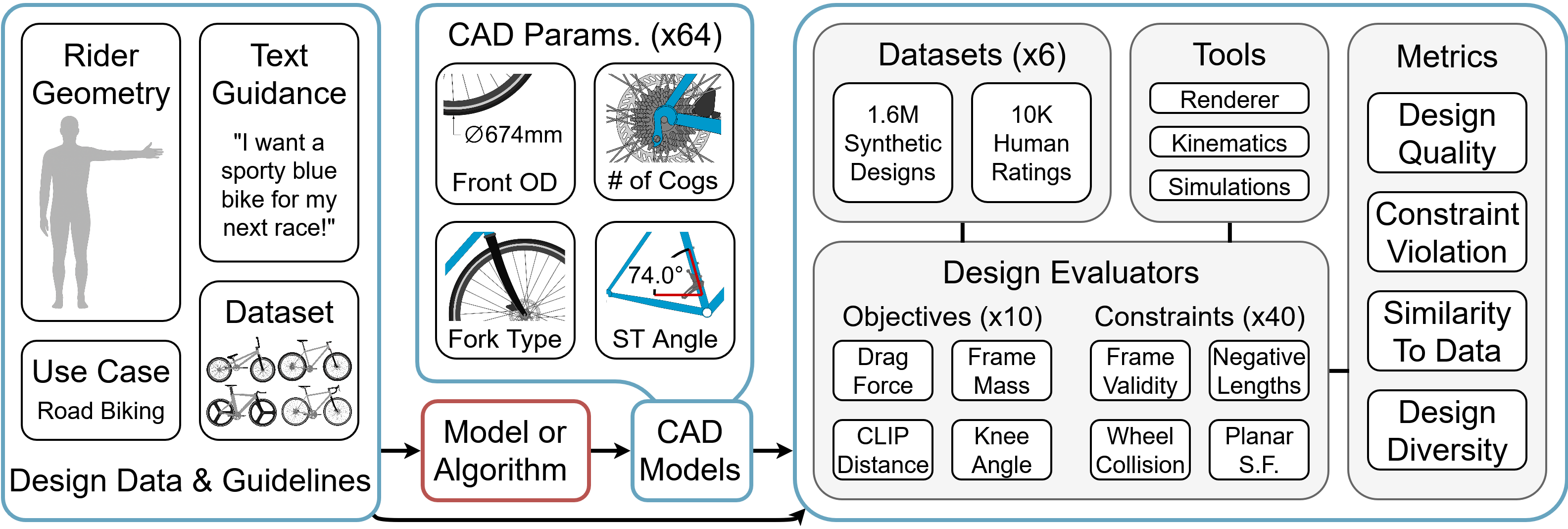}
    \caption{High-level overview of the benchmarking procedure.}
    \label{fig:procedure}
\end{figure}

The design objectives detailed above yield a set of 10 design objectives and 40 design constraints, which are explained in detail in the Appendix~\ref{appx:objlist}. However, to compare the performance of different generative models, we desire a concise set of summary metrics to evaluate sample sets. BikeBench's metrics and general benchmarking procedure are introduced in Figure~\ref{fig:procedure}, and described in detail in this section.

\subsection{Metrics} Though there are countless possibilities for generative model evaluation metrics~\cite{regenwetter2023beyond}, we select four principal metrics: design quality, constraint violation, similarity to data, and design diversity, which we describe below. Unlike other benchmarks with singular objectives, models benchmarked on BikeBench should be compared across all four scoring metrics. We also evaluate a few auxiliary bonus objectives, as discussed in Appendix~\ref{appx:metrics}.

\paragraph{Design Quality (Hypervolume Metric):}
High-quality design tools should generate high-quality designs. To quantify design quality, we calculate the hypervolume metric over any designs that simultaneously satisfy all constraints. The hypervolume metric is a staple of multi-objective optimization literature which calculates the overall multi-objective optimality of a set of designs. Because hypervolume directly benefits from constraint satisfaction and diversity, hypervolume can be seen as the best single metric to indicate design utility. More details on the hypervolume metric and its calculation are included in the Appendix~\ref{appx:metrics}. 

\paragraph{Constraint Violation:}
Effective design synthesis tools can reliably navigate design constraints. We measure the mean number of constraints that the model violates per design. 

\paragraph{Similarity to Data (Maximum Mean Discrepancy):} 
Basing new designs on existing ones can be an effective method to minimize risk and maximize feasibility. To capture a model's ability to match the manifold of existing designs, we evaluate distributional similarity to a subset of 759 designs withheld from the model during training. We select Maximum Mean Discrepancy, a common kernel-based statistical discrepancy measure used to compare sets of samples. Statistical similarity serves to enforce two desirable attributes in generative models for design~\cite{regenwetter2023beyond}. The first is `design realism'---what does it mean to be a bike and not something entirely different, say, a wheelbarrow? The second desirable attribute is `design space coverage,' encouraging models to learn more than just a small, niche subspace of the overall breadth of bicycle designs, expanding their utility.

\paragraph{Design Diversity (DPP Diversity):} Tools that create a diverse set of design solutions offer the most selection to human users.  To capture the diversity of generated designs, we measure a Determinantal Point Process (DPP)-based diversity score, as described in~\cite{regenwetter2023beyond}. More details on DPP diversity are included in Appendix~\ref{appx:oagm_training}.

\subsection{Benchmarking Procedure}
BikeBench's benchmarking procedure evaluates 100 designs for each of 100 fixed conditional cases (text prompt, rider anthropometric dimensions, use case). Distributional similarity, constraint satisfaction rate over all 40 constraints, and hypervolume metric over all 10 evaluation objectives are evaluated for each set of 100 bikes, and the mean over the 100 conditions is reported as the final summary score for each metric. 

To avoid bias, models are trained only on the 3036 training samples from the original 3798 human-designed bicycles, and not on synthetically generated designs (the synthetic dataset is used exclusively for evaluation of aesthetics scores). Models are allowed to train on either a mixed-datatype or continuous version of the data. Designs generated by models trained on continuous data are mapped back to a mixed-datatype representation prior to evaluation---booleans are decided using rounding, while one-hot categorical data is decided using argmax. 

\subsection{Benchmark Settings}
BikeBench is configurable to provide benchmarking scenarios that accommodate a variety of real-world application cases. This configurability is described below:
\paragraph{Conditional Generation:} In the conditional generation setting, the test-set conditions can only be accessed after all evaluator calls are complete, forcing the method to have conditional generation capabilities (zero-shot generation on new conditions). 
\paragraph{Gradient-Free Evaluation:} Although all of BikeBench's evaluators offer gradients, gradients can be turned off to benchmark gradient-free methods. 
\paragraph{Masked Constraints:} To assess models' ability to infer constraints implicitly from data, we offer a benchmarking case where all but six of the 32 geometric validity constraints are masked to the model. The six retained are the constraints that are violated by at least 5\% of the dataset, where the dataset wouldn't be a good baseline for inferring them. 

\paragraph{Evaluation Limits:} Since design evaluation can be costly, design synthesis methods typically try to minimize the number of evaluation calls. BikeBench tallies the number of design evaluations used by each model. Note that for unconditional methods like optimization, evaluations will be summed across all test conditions. To standardize methods, we provide a few max-evaluation brackets: 0, 1K, 1M, 1B, and unlimited. 

\begin{figure}[!htb]
    \centering
    \includegraphics[width=\linewidth]{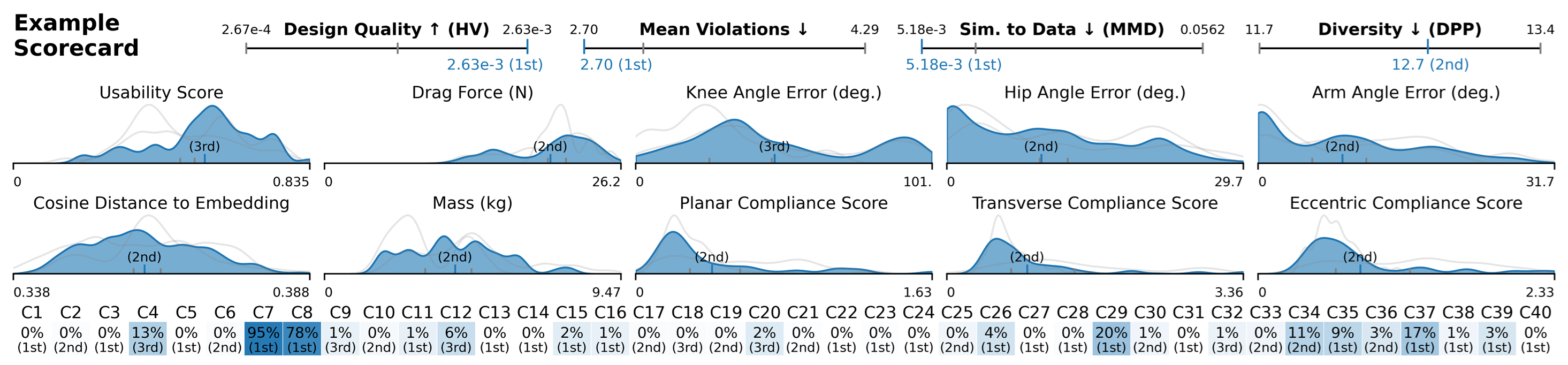}
    \caption{Example scorecard. The scorecard shows key model performance metrics at the top, objective score distributions in the middle, and constraint violation rates at the bottom.}
    \label{fig:scorecard}
\end{figure}

We establish the ``standard'' BikeBench settings as follows: Models can call at most 1M gradient-supporting evaluations and will receive completely unmasked scores. However, our leaderboard will support easy filtering of methods and a few preset alternative configurations so that users can easily compare and report scores on non-standard tasks. We request that users who make comparative performance claims on non-standard benchmarks clearly report the BikeBench settings under which they are making these claims. Our benchmark also includes standardized scorecards that showcase key model scores, objective distributions, and constraint satisfaction rates, as shown in Figure~\ref{fig:scorecard}.

\section{Models Benchmarked}
Our principal goal in the benchmarking of baselines is to span a variety of types of design generation procedures ranging from LLMs to tabular generative models to optimization algorithms. Because we do not focus on extensively comparing methods within specific classes, we limit our benchmarking to just one or a few methods per class. The various features and attributes of the models tested are summarized in Appendix~\ref{appx:modelsandmetrics}. 
\paragraph{Baselines:} Our benchmarking starts with two simple baselines. The first is pure random sampling from the dataset, which is completely agnostic to design quality. The second is rejection sampling from the dataset based on evaluated validity scores. 
\paragraph{Large Language Models:}
To gauge the general performance of Large Language Models on BikeBench, we provide an LLM interface with prompts, text descriptions of design parameters and objectives, and a data loader, as described in Appendix~\ref{appx:LLMdetails}. We benchmarked OpenAI's GPT-5-high. At the time of benchmarking (October 20, 2025) this was the highest performing model on ArtificialAnalysis's LLM intelligence index, a blended evaluation suite of seven individual evaluations~\cite{wang2024mmlu, phan2025humanity, rein2024gpqa, hendrycks2021measuring, tian2024scicode, jain2024livecodebench}. To comply with the model's context window, we exposed it to examples of 25 valid designs and 25 invalid designs, their associated conditions, and their corresponding evaluation scores resulting from the condition–design pairs. The model was then provided one new condition and asked to generate a CSV file with the 100 solution designs. This was repeated for the 100 conditions. In the conditional case, the examples provided are for the same condition that the model is generating for. 

\paragraph{Tabular Generative Models:}
BikeBench, like many tabular data problems, contains ordinal, boolean, and categorical features. Training deep learning models directly on categorical data generally requires special modifications, particularly in categorical data generation settings where gradients must be propagated through this discretization. Several generative modeling formulations have been proposed to generate mixed continuous and categorical data. We benchmark two popular models, CTGAN and TVAE, proposed in~\cite{xu2019modeling}. These models do not consider design performance, and are benchmarked principally for their distribution-learning capabilities. Since these models do not utilize the design evaluation budget during training, the budget is spent on rejection sampling of generated designs. 

\paragraph{Optimization-Augmented Generative Models:}
Performance-augmented training is a popular approach to train generative AI models to synthesize high-quality designs. One method involves calling the design evaluation function during the training process, effectively realizing an optimization-augmented generative model. To balance a tendency toward mode collapse, a diversity-based auxiliary loss can encourage diverse generation of high-performing designs~\cite{chen2021padgan}. This technique has subsequently been extended to handle multiple objectives and constraints~\cite{regenwetter2022design}. We benchmark this optimization-augmented training formulation using a GAN, VAE, and DDPM as base models. We also test a DDPM variant where the auxiliary objective is instead applied during guidance (labeled with~\texttt{-G}). All optimization-augmented benchmarks are labeled with the~\texttt{OA-} prefix in our benchmarks. More details on our implementation are included in Appendix~\ref{appx:oagm_training}. 

\paragraph{Optimization:} 
We benchmark both gradient-based and heuristic optimization. We test plain aggregation-based gradient descent~\cite{miettinen1999nonlinear}, implemented in the LibMoon optimization library~\cite{zhang2024libmoon}. Gradient descent is applied to one-hot encoded data with constraints enforced using a 1000x weighted penalization function. We also benchmark a mixed-variable adaptation of the Non-dominated Sorting Genetic Algorithm~II (NSGA-II)~\cite{deb2002fast}, a staple evolutionary optimization algorithm, as implemented in pymoo~\cite{blank2020pymoo}. Whereas the gradient-based algorithms benefit from the efficiency of gradient information, the mixed-variable NSGA-II operates directly on the mixed-variable data without obfuscation caused by one-hot encoding or continuous relaxation of constraints. All optimization algorithms are run once for each of the 100 test conditions, splitting the evaluation budget. Optimization is initialized per-condition using valid designs from the dataset. 

\section{Benchmarking Results}
In this section, we present and analyze the benchmarking results using the aforementioned ``standard'' evaluation procedure: unmasked constraints, gradients, and a 1M evaluation budget. Results in both unconditional and conditional generation are compiled in Table~\ref{tab:unmasked_all}. All models are evaluated on design quality, constraint violation, similarity to dataset, and diversity. Detailed scorecards for all tested models are included in Appendix~\ref{appx:scorecards}.

\begin{table}[!htb]
\centering
\caption{Design quality, constraint violation, similarity to dataset, and design diversity scores for unconditional and conditional benchmarking cases. Models are separated by class (LLM / Tabular-GM / OA-GM / Optimizer). The best non-baseline scores in each metric are bolded. Models are benchmarked on unmasked constraints with an evaluation budget of 1M.}
\label{tab:unmasked_all}
\resizebox{\textwidth}{!}{%
\begin{tabular}{lcccccccc}
\toprule
 & \multicolumn{4}{c}{Unconditional Generation} & \multicolumn{4}{c}{Conditional Generation} \\
\cmidrule(lr){2-5} \cmidrule(lr){6-9}
 & Qual. ($\uparrow$) & Viol. ($\downarrow$) & Sim. ($\downarrow$) & Div. ($\downarrow$) &
   Qual. ($\uparrow$) & Viol. ($\downarrow$) & Sim. ($\downarrow$) & Div. ($\downarrow$) \\
\midrule
Dataset      & 0.0035 & 2.699 & 0.005 & 12.72 & 0.0035 & 2.699 & 0.005 & 12.72 \\
Data subset  & 0.0151 & 0.000 & 0.081 & 12.28 & 0.0151 & 0.112 & 0.079 & 12.43 \\
\midrule
GPT-5-high            & 0.0086 & 0.448 & 0.363 & 14.08 & 0.0049 & 1.288 & 0.631 & 14.70 \\
\midrule
CTGAN                 & 0.0087 & \textbf{0.000} & 0.147 & \textbf{7.10} & 0.0148 & 0.318 & 0.082 & \textbf{11.93} \\
TVAE                  & 0.0160 & \textbf{0.000} & 0.163 & 13.55 & 0.0161 & 0.053 & 0.153 & 13.55 \\
\midrule
OA-GAN                & 0.0137 & \textbf{0.000} & 0.308 & 14.44 & 0.0109 & 2.579 & \textbf{0.079} & 14.04 \\
OA-VAE                & 0.0181 & \textbf{0.000} & 0.301 & 14.96 & \textbf{0.0186} & \textbf{0.001} & 0.348 & 14.97 \\
OA-DDPM               & 0.0168 & \textbf{0.000} & 0.077 & 12.55 & 0.0165 & 0.042 & 0.084 & 12.82 \\
OA-DDPM-G        & \textbf{0.0195} & \textbf{0.000} & \textbf{0.037} & 13.03 & - & - & - & - \\
\midrule
Grad. Descent         & 0.0164 & 0.092 & 0.266 & 12.04 & - & - & - & - \\
NSGA-II               & 0.0102 & \textbf{0.000} & 0.622 & 11.88 & - & - & - & - \\
\bottomrule
\end{tabular}%
}
\end{table}

\subsection{Discussion}

\paragraph{Baselines:} The relatively high constraint violation of the dataset baseline (avg. 2.7 constraints per design) may initially seem surprising. The structural safety factor is a key driver of this low validity: $95\%$ and $80\%$ of dataset designs fail the planar and eccentric safety factor constraints, respectively. This systematic under-engineering is one of the main sources of bias in the BIKED dataset, arising from the fact that many designers never adjust tube thickness parameters, since they are not visually prominent in the BikeCAD user interface~\cite{regenwetter2023framed}. This presents an interesting test for generative models, which, to satisfy structural constraints, generally have to deviate from the norms of the dataset to systematically thicken certain tubes. This exemplifies the principal challenge of BikeBench: \textbf{To simultaneously succeed across metrics, models must subtly but strategically deviate from the distribution of the dataset to achieve design goals.} Although sampling only valid designs from the dataset is a fairly strong baseline, strong models should be able to outperform it.

\paragraph{Large Language Model:}
The rich history of bicycle design has extensive textual documentation and almost certainly appears in common LLM training datasets. Consequently, LLMs can theoretically leverage contextual knowledge of the design parameters and the evaluation criteria to gain an edge over other classes of models. However, they must overcome key disadvantages in their lack of specialization for tabular data, as well as limited context windows, which may prevent them from observing the full dataset. OpenAI's GPT-5-high, the lone LLM tested, was a clear underperformer across all metrics, ranking in the bottom three methods for each objective. This indicates, unsurprisingly, that LLMs have substantial room for growth in constrained engineering design problems and tabular-data domains. 

\paragraph{Tabular Generative Models:}
The tabular generative models trained were unconditional and agnostic to design constraints and objectives. The evaluations were only invoked to sample subsets of generated designs. Across the board, their design quality, constraint violation, and similarity to data were generally equal to or worse than the simple baseline of dataset subsampling. However, CTGAN was a notable frontrunner in diversity score. These benchmarking results highlight the importance of performance-aware training, compared to subsampling of pure performance-agnostic generated samples. 

\paragraph{Optimization-Augmented Generative Models:}
Optimization-Augmented Generative Models achieved the best quality, constraint violation, and similarity scores in both unconditional and conditional generation. While the guided DDPM was especially dominant in the unconditional setting, the standard DDPM was a strong performer across the board. The VAE was a strong performer in quality but its similarity and diversity scores were much worse than the DDPMs. Finally, the GAN underperformed its optimization-augmented counterparts in most metrics. In general, the OA-GMs achieved strong balanced scores with standout performance from optimization-augmented DDPM models. 

\paragraph{Optimization:}
Classic optimization algorithms were relatively unimpressive performers on BikeBench. However, their diversity scores were notably strong, indicating a capability to maintain a wider variety of high-quality solutions compared to competing methods. Interestingly, the diversity of AI-generated design sets is often one of their selling points over classically-optimized design sets. BikeBench's results indicate that excessive fixation on design quality can cause AI-generated designs to be less diverse than classically-optimized designs. Gradient-based optimization achieved reasonably strong design quality scores with decent similarity. NSGA-II struggled in both design quality and dataset similarity. The performance of these classic optimization algorithms highlights a major shortcoming: Even when amortizing evaluation budget over just 100 designs, generative models outperform optimization in classic optimization metrics. This suggests considerable room for growth, perhaps using approaches that would amortize learning across conditional cases.

\section{Conclusion and Future Directions}
We proposed BikeBench, a constrained engineering design benchmark for generative models, comprising 10 multiphysics and human-centered objectives and 40 geometric, ergonomic, and safety constraints. We benchmarked a variety of design generation procedures, including an LLM, tabular generative models, optimization algorithms, and optimization-augmented generative models. Our benchmarking results suggest optimization-augmented models achieve well-rounded performance, optimization and tabular generative models achieve mixed results, and LLMs generally underperform. 

We encourage further benchmarks of generative models, optimization algorithms, and design generation procedures that transcend boundaries. LLMs with larger context windows or tabular data adapters, foundation models for tabular generation, cross-conditional amortization of optimization, and optimization-augmented, mixed-variable models may be of particular interest. Lastly, we advocate for the development of more multimodal engineering design benchmarks of complex systems. Alongside BikeBench, such benchmarks will help expand the frontier of generative AI in engineering design and beyond.

\begin{ack}
We thank Amin Heyrani Nobari for his contributions to the renderer and Ankur Kapileshwar for his contributions to the geometric validation functions. We also acknowledge support from the MIT MISTI–Netherlands Program.
\end{ack}

\bibliographystyle{IEEEtran}
\bibliography{biblio}

\newpage
\appendix

\section{Appendix A: Extended Results} 
\subsection{Scorecards for Main Paper Results} \label{appx:scorecards}
\begin{figure}[h]
    \centering
    \includegraphics[width=\linewidth]{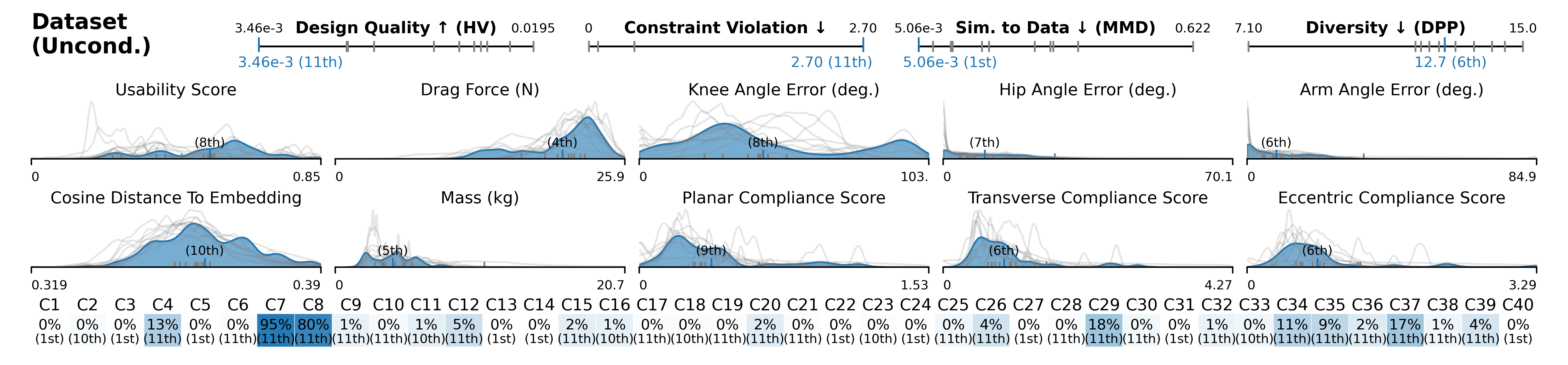}
    \includegraphics[width=\linewidth]{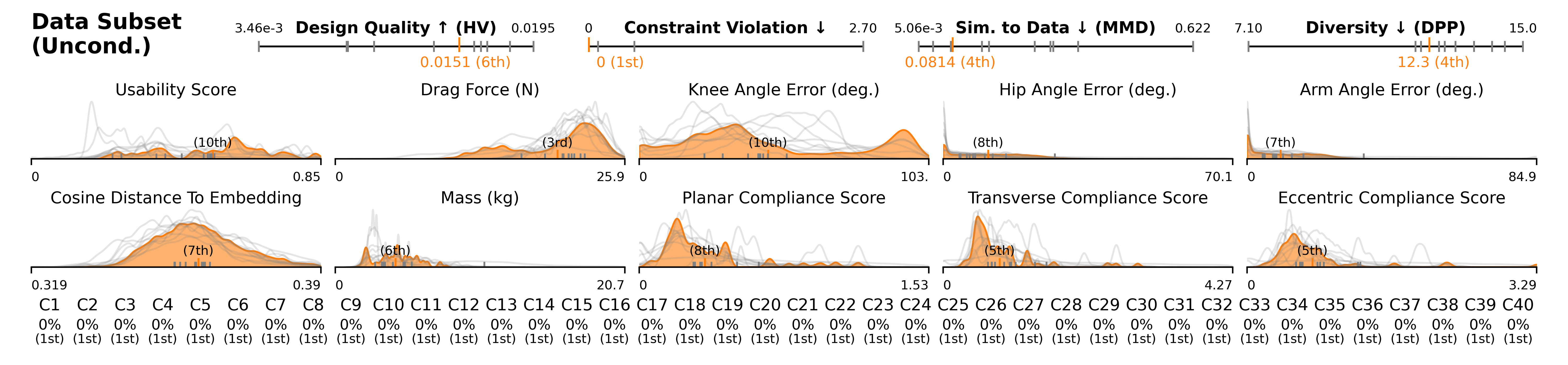}
    \includegraphics[width=\linewidth]{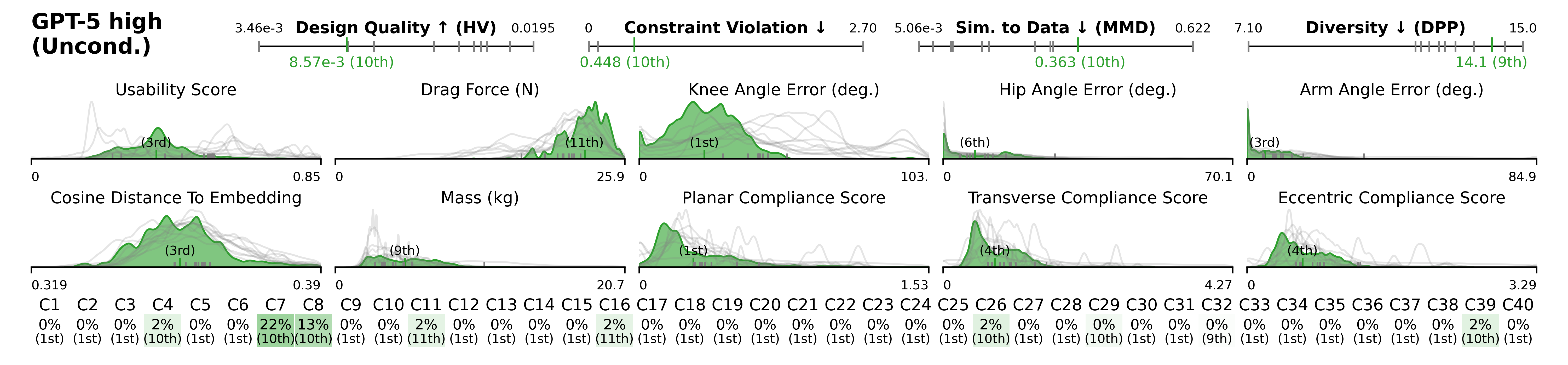}
    \includegraphics[width=\linewidth]{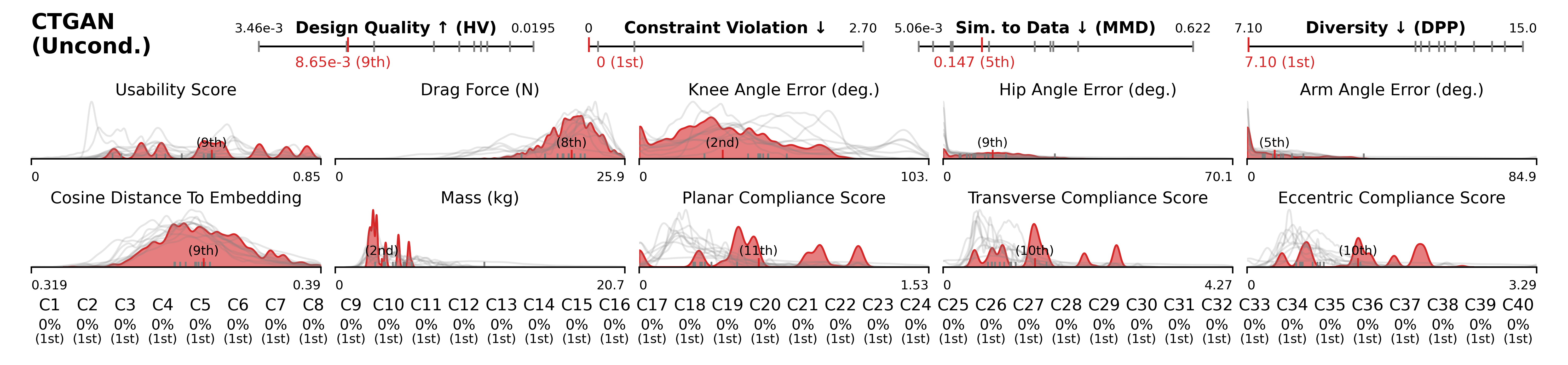}
    \includegraphics[width=\linewidth]{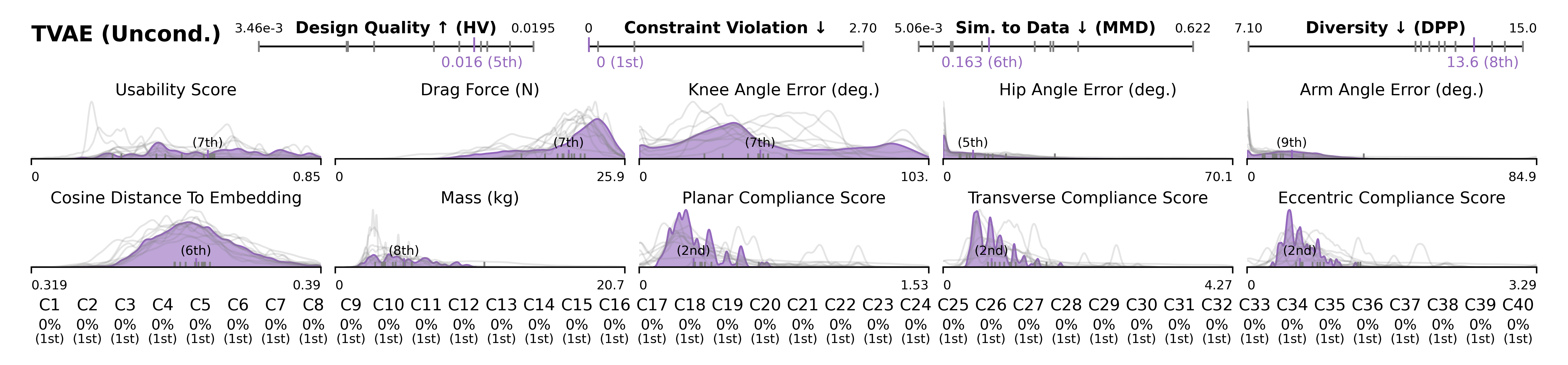}
    \caption{Scorecards for unconditional benchmarking results (part 1)}
    \label{fig:unconditional1}
\end{figure}

\begin{figure}[h]
    \centering
    \includegraphics[width=\linewidth]{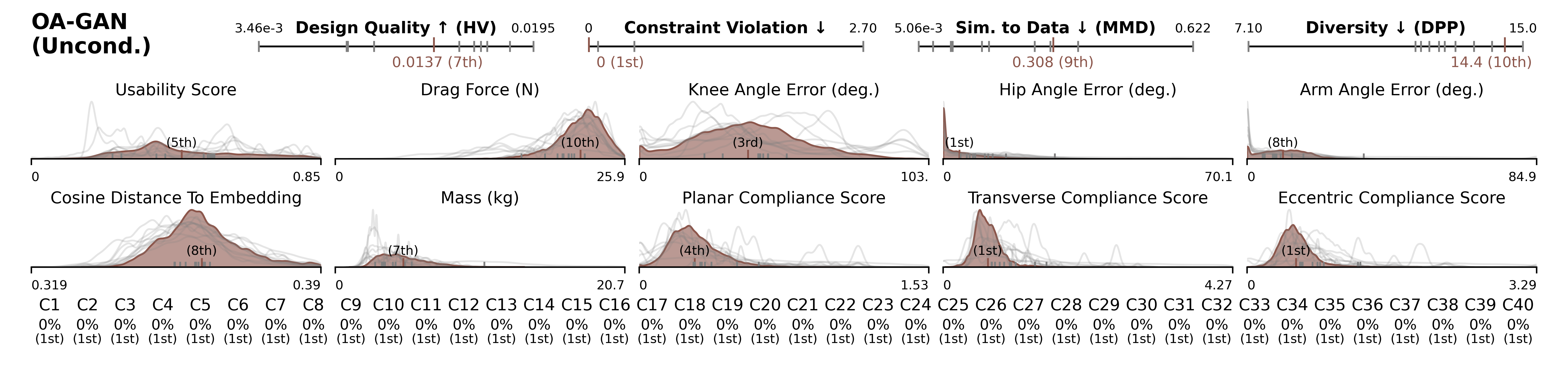}
    \includegraphics[width=\linewidth]{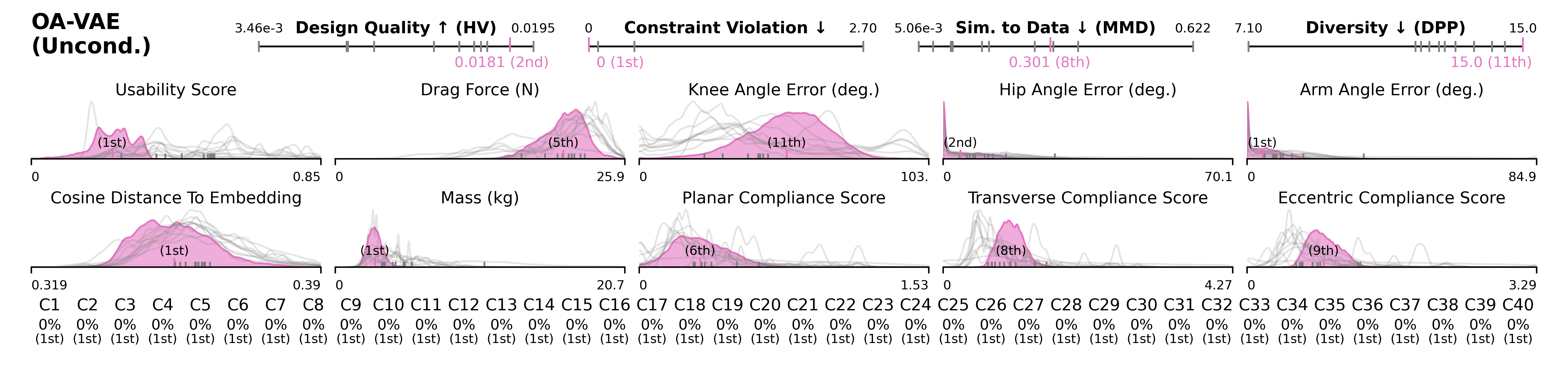}
    \includegraphics[width=\linewidth]{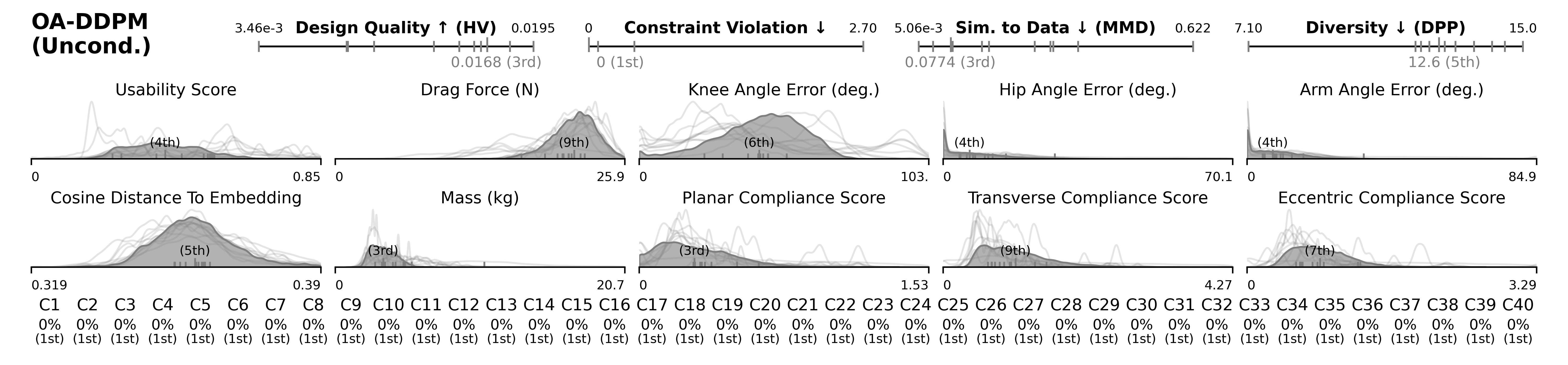}
    \includegraphics[width=\linewidth]{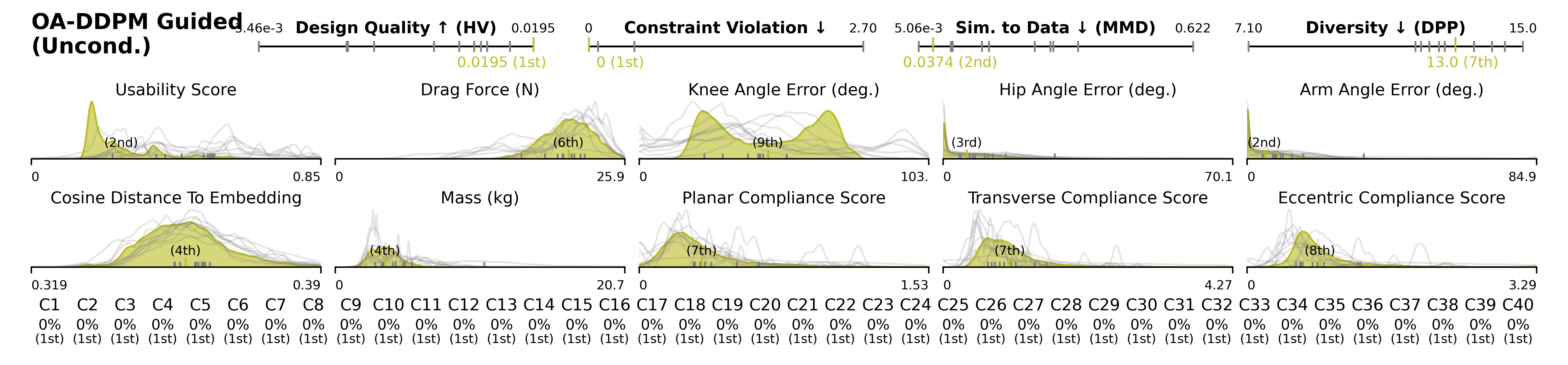}
    \includegraphics[width=\linewidth]{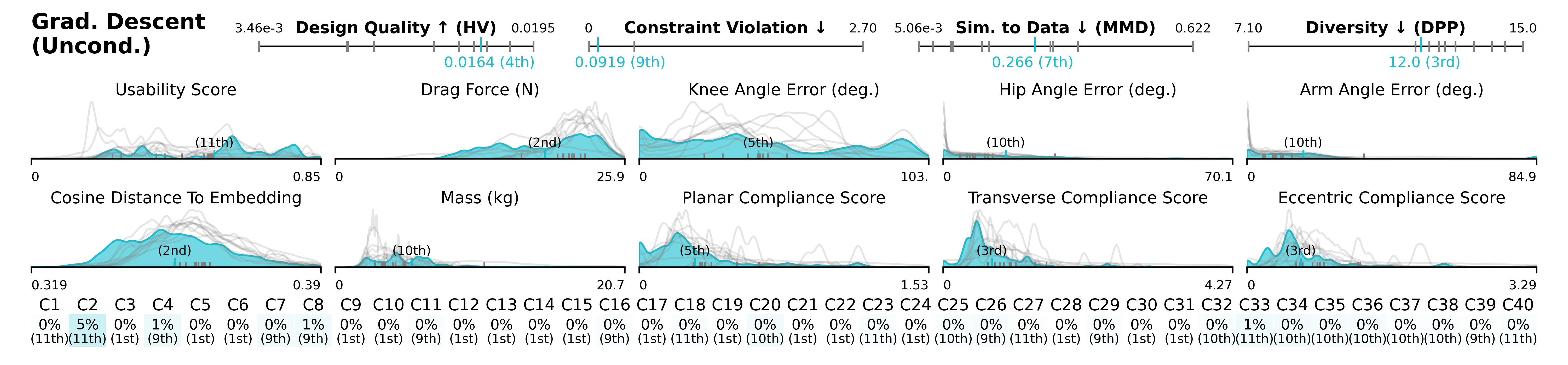}
    \includegraphics[width=\linewidth]{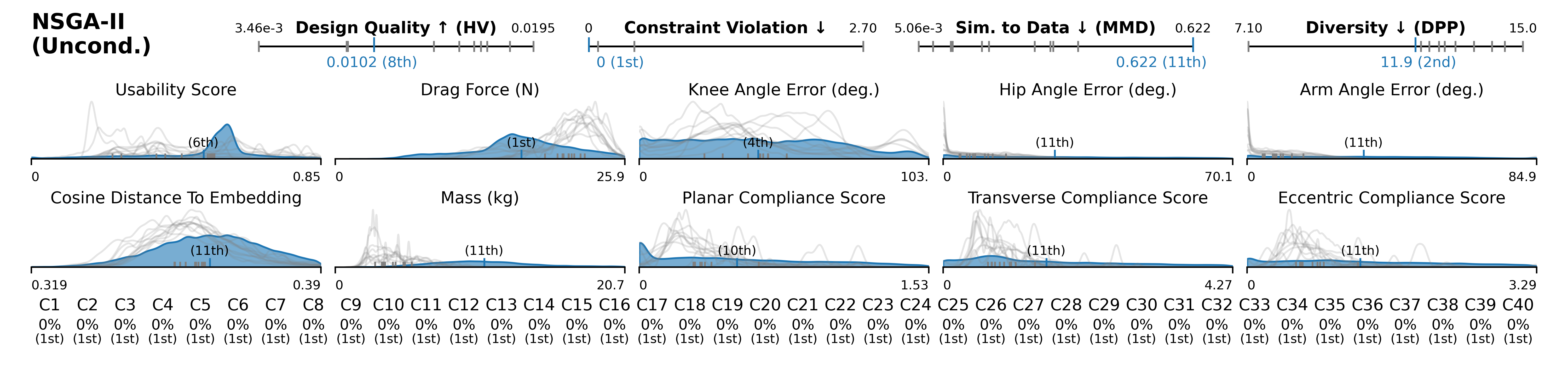}
    \caption{Scorecards for unconditional benchmarking results (part 2)}
    \label{fig:unconditional2}
\end{figure}

\begin{figure}[h]
    \centering
    \includegraphics[width=\linewidth]{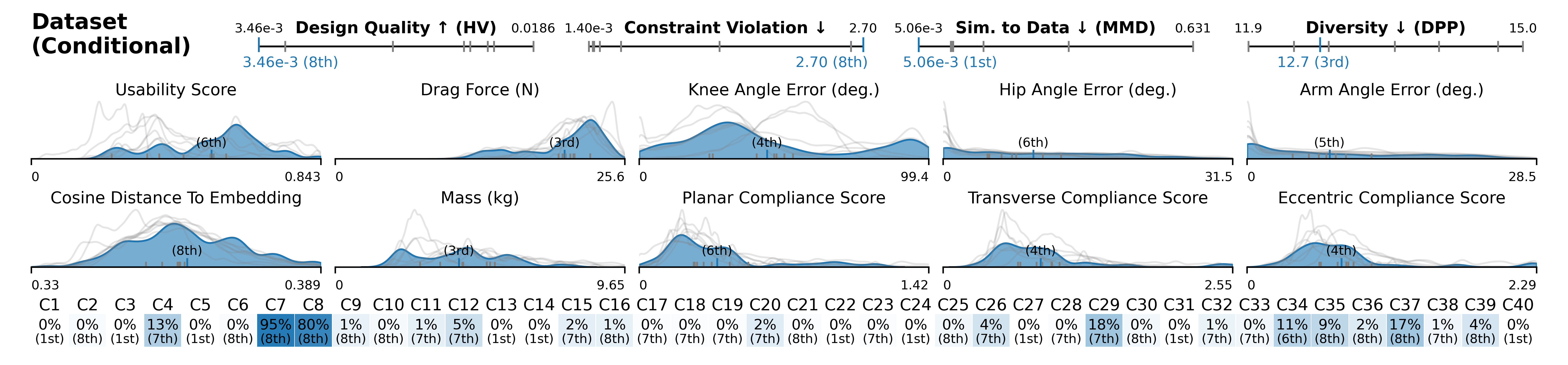}
    \includegraphics[width=\linewidth]{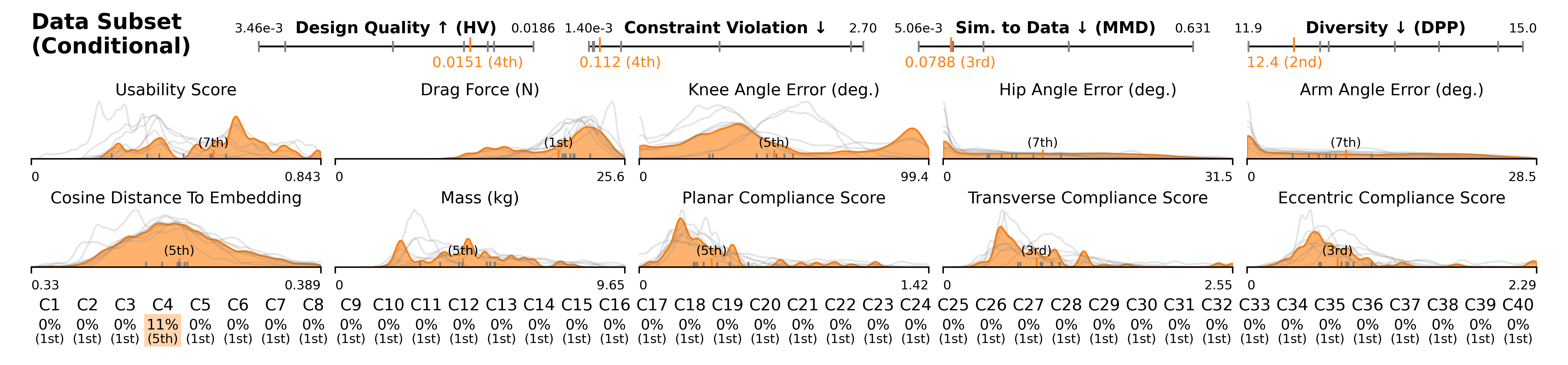}
    \includegraphics[width=\linewidth]{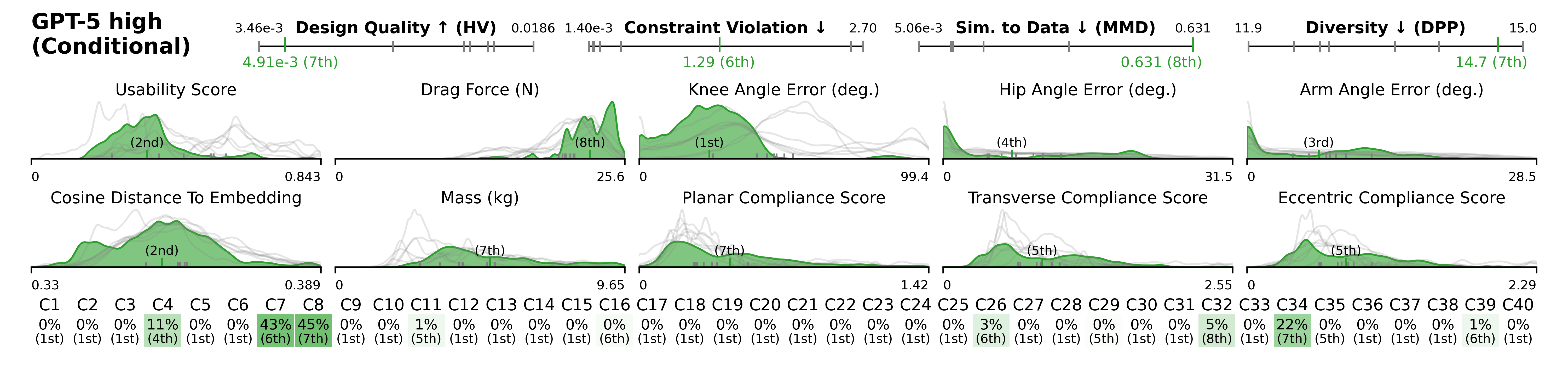}
    \includegraphics[width=\linewidth]{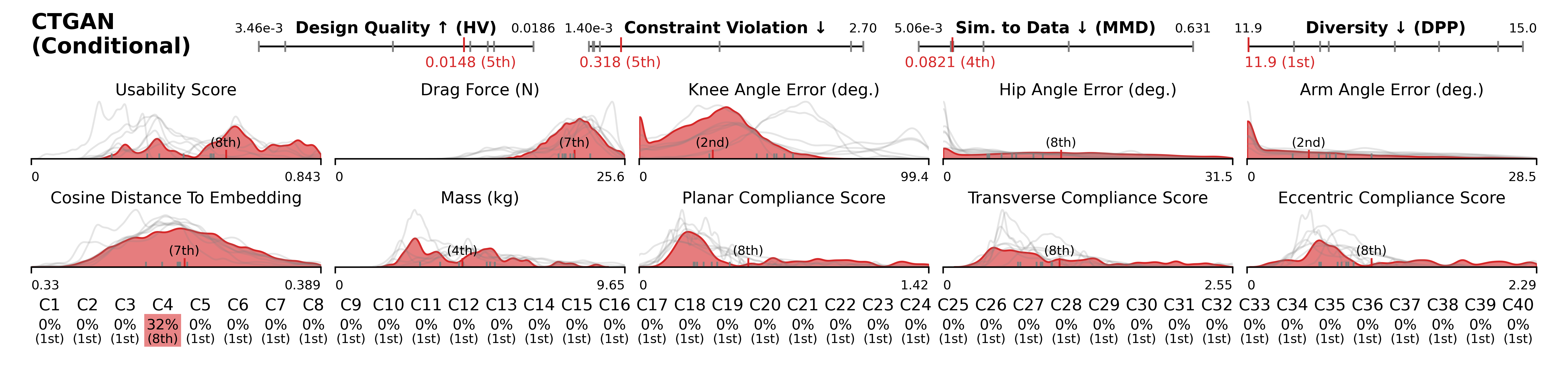}
    \includegraphics[width=\linewidth]{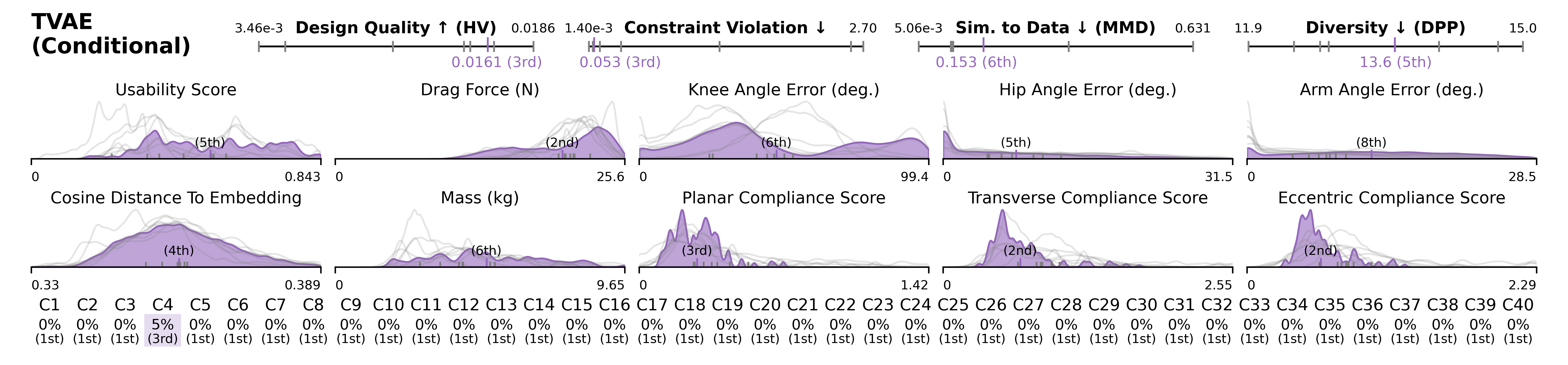}
    \includegraphics[width=\linewidth]{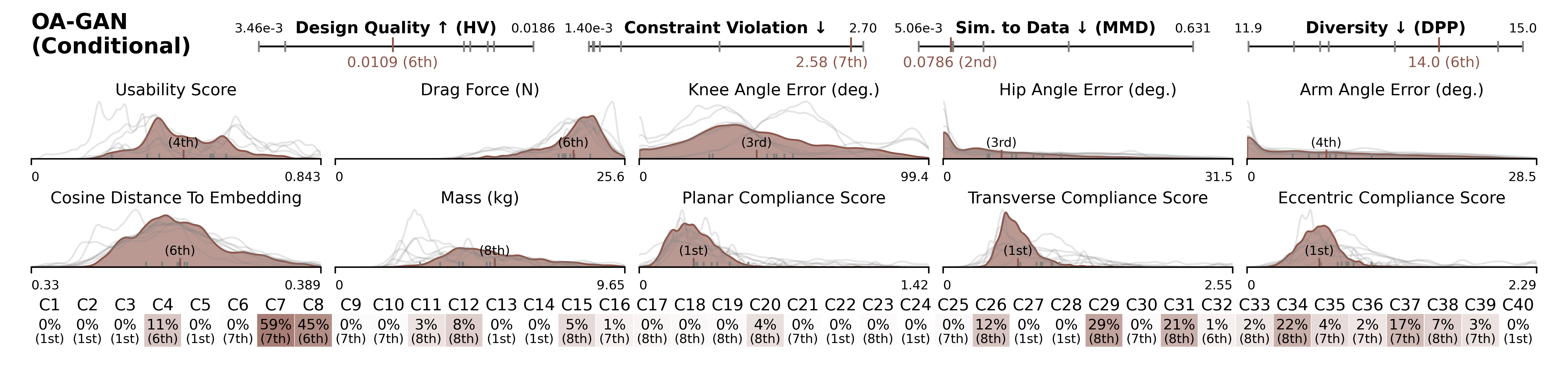}
    \caption{Scorecards for conditional benchmarking results (part 1)}
    \label{fig:conditional1}
\end{figure}

\begin{figure}[h]
    \centering
    \includegraphics[width=\linewidth]{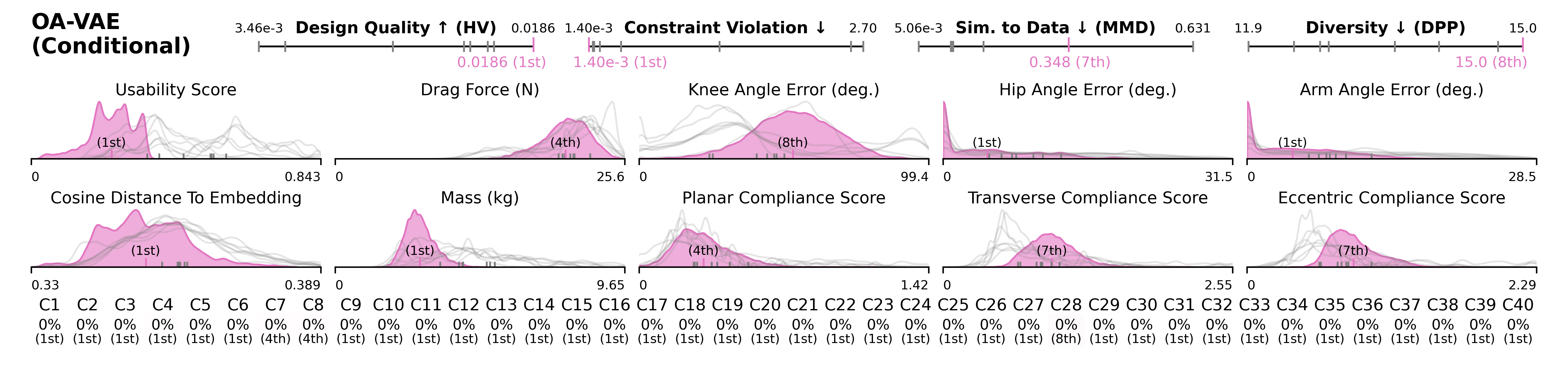}
    \includegraphics[width=\linewidth]{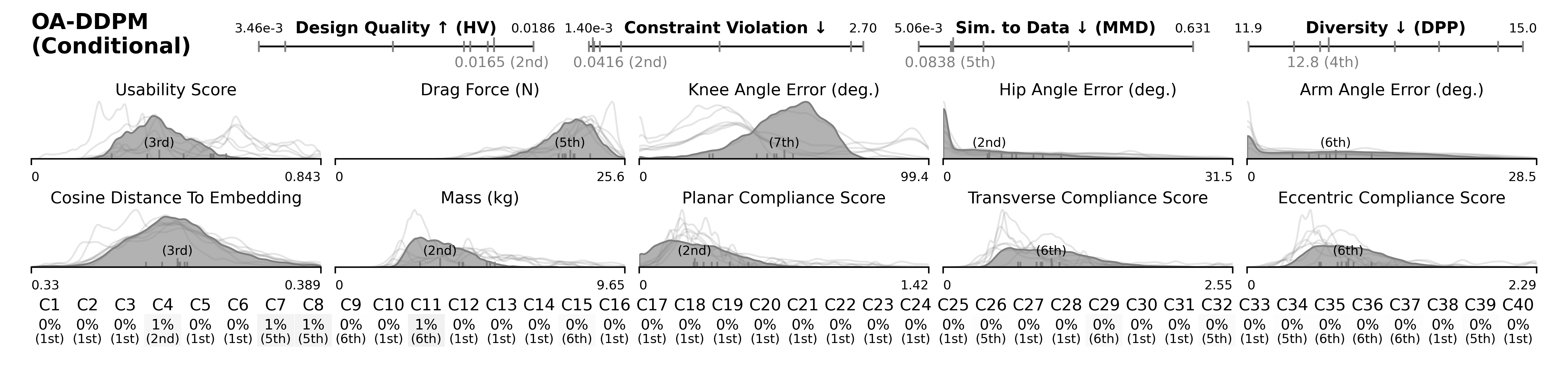}
    \caption{Scorecards for conditional benchmarking results (part 2)}
    \label{fig:conditional2}
\end{figure}

We include scorecards for all models benchmarked in the full paper. First, we show unconditional generation in Figures~\ref{fig:unconditional1} and~\ref{fig:unconditional2}. Next, we include conditional generation scorecards in Figures~\ref{fig:conditional1} and~\ref{fig:conditional2}. 
\clearpage

\clearpage

\clearpage

\subsection{Generation Under Masked Constraints}
We include benchmarking results for the masked constraint case under both conditional and unconditional generation in Table~\ref{tab:masked-results}. GPT-5-high is excluded here due to time and budgetary constraints. In general, models with lower similarity scores struggle more to infer constraints from data. These models therefore tend to suffer a larger reduction in design quality and increase in constraint violation compared to the unmasked generation case. Few methods compete with the simple dataset subsampling baseline. This broadly suggests significant room for improvement of design synthesis algorithms in implicit constraint satisfaction. 

\begin{table}[!htb]
\centering
\caption{Design quality, constraint violation, similarity to dataset, and design diversity scores for unconditional and conditional benchmarking cases with \textbf{masked constraints}. Models are separated by class (LLM / Tabular-GM / OA-GM / Optimizer). The best non-baseline scores in each metric are bolded. Models are benchmarked with an evaluation budget of 1M.}
\label{tab:masked-results}
\resizebox{\textwidth}{!}{%
\begin{tabular}{lcccccccc}
\toprule
 & \multicolumn{4}{c}{Masked Generation} & \multicolumn{4}{c}{Conditional + Masked Generation} \\
\cmidrule(lr){2-5} \cmidrule(lr){6-9}
 & Qual. ($\uparrow$) & Viol. ($\downarrow$) & Sim. ($\downarrow$) & Div. ($\downarrow$) &
   Qual. ($\uparrow$) & Viol. ($\downarrow$) & Sim. ($\downarrow$) & Div. ($\downarrow$) \\
\midrule
Dataset      & 0.0035 & 2.699 & 0.005 & 12.72 & 0.0035 & 2.699 & 0.005 & 12.72 \\
Data subset  & 0.0151 & 0.123 & 0.082 & 12.41 & 0.0151 & 0.216 & 0.076 & 12.51 \\
\midrule
CTGAN                 & 0.0079 & 1.046 & 0.123 & \textbf{9.94} & 0.0109 & 1.301 & \textbf{0.098} & \textbf{11.65} \\
TVAE                  & 0.0150 & 0.616 & 0.164 & 13.55 & 0.0151 & \textbf{0.428} & 0.154 & 13.54 \\
\midrule
OA-GAN                & 0.0128 & \textbf{0.505} & 0.196 & 14.02 & 0.0111 & 2.390 & 0.105 & 14.12 \\
OA-VAE                & 0.0142 & 0.971 & 0.341 & 14.99 & \textbf{0.0157} & 1.062 & 0.284 & 14.92 \\
OA-DDPM               & 0.0087 & 1.541 & \textbf{0.101} & 11.06 & 0.0007 & 2.495 & 0.201 & 12.93 \\
\midrule
Grad. Descent         & \textbf{0.0161} & 0.683 & 0.279 & 12.09 & - & - & - & - \\
NSGA-II               & 0.0011 & 3.566 & 0.606 & 11.59 & - & - & - & - \\
\bottomrule
\end{tabular}%
}
\end{table}

\clearpage
\section{Details on Background, Datasets, Evaluation Criteria, Metrics, and Models}
\subsection{Background: Data-Driven Bicycle Design} 
\label{appx:bikedesignbackground}
Bicycle design is a complex engineering problem with a rich history of optimization through scientific and engineering innovation~\cite{sharp1896bicycles}. Bicycle design catalyzed numerous significant advancements in mechanical engineering and remains a significant research field to this day. Bicycles themselves revolutionized transportation and remain ubiquitous in today's society. In 2015, at least 42\% of households globally owned a bike~\cite{oke2015tracking} and 35\% of adults surveyed across 28 countries in 2022 rode a bike at least once a week~\cite{ipsos2022cycling}. Thanks to this history, the widespread continued use, and the hugely varied use cases and subjective preferences among users, the bicycle design space is rich and vibrant -- ideal for data-driven methods.

Data and computation have played an increasingly important role in bicycle design science~\cite{de1999quantification}, predominantly focusing on computational simulation and optimization~\cite{lessard1995utilization, godo2009aerodynamic,covill2016assessment}. However, little published research had explored bicycle design-space exploration or big-data applications before the BIKED~\cite{regenwetter2022biked} dataset was released in 2021. BIKED demonstrated some of the first applications of design-space modeling, deep learning, and generative AI in bicycle design. 

Data-driven tools are well-positioned to continue the legacy of innovation in bicycle design science. Indeed, data-driven design tools stemming from BIKED have already been integrated into professional bicycle design software (see www.bikecad.ca/ai). Thanks to the continued ubiquity of the bicycle in modern society, design innovation may yield more optimal or better-customized bikes, potentially increasing ridership. Such an increase in ridership could subsequently impart further societal impact by improving public health~\cite{oja2011health}, traffic congestion~\cite{hamilton2018bicycle}, and climate change~\cite{edenhofer2015climate}. 

\subsection{Datasets and Evaluation Criteria}
\subsubsection{Details on Data Collection from Human Subjects} \label{appx:humansubject}
All participants in our human subject data collection were adults (aged 18 or older), paid an hourly rate of 9 GBP. All participants were informed that their data may be used for the training and evaluation of ML/AI models and consented to this. The study passed relevant institutional review procedures without issue.

\subsubsection{Building a Classification Dataset for Usability} \label{appx:usabilityclassification}

\begin{figure}[!htb]
    \centering
    \includegraphics[width=0.65\linewidth]{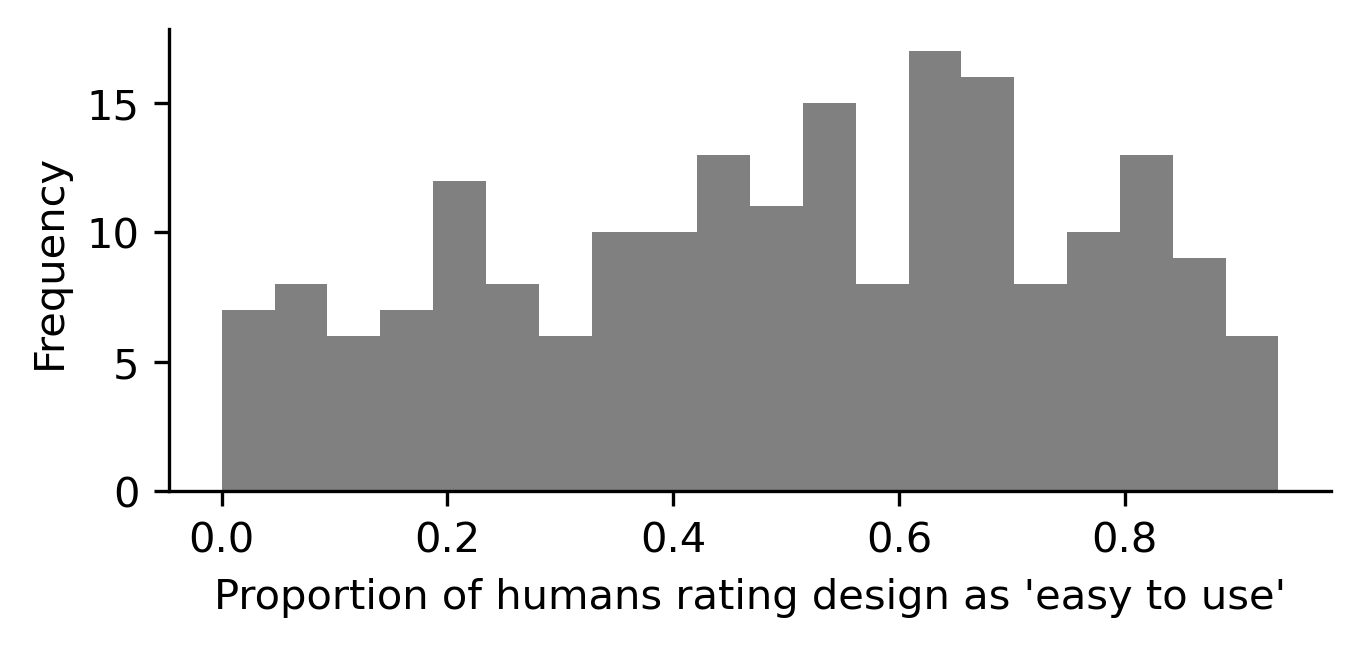}
    \caption{Distribution of average usability scores in BikeBench's human-sourced dataset.}
    \label{fig:usabilityhist}
\end{figure}

BikeBench uses the binary (yes/no) ratings collected from human raters to predict the proportion of raters that would consider a design `easy to use.' Rather than averaging the scores for each bike, yielding the distribution of scores shown in Figure~\ref{fig:usabilityhist}, ratings can alternatively be aggregated to form a simple classification dataset. To assess the overall consensus, factoring in a margin of error resulting from potential sampling inaccuracies, we can use a two-sided binomial test. Taking the population size of the group with the smallest number of valid users (46), we evaluate the $99\%$ confidence interval to be approximately \textbf{20\%}. Thus, for any design with at least 70\% consensus among respondents, we can be 99\% confident in the consensus value. In other words, an average score of at least 0.7 confidently indicates that more than half of raters would consider the bike usable, whereas an average of at most 0.3 indicates that less than half of raters would consider the bike usable. Following these thresholds, \textbf{49} bikes are identified as usable and \textbf{51} as unusable, resulting in a total of \textbf{100 confident classification labels}. 

\subsubsection{Limitations and Assumptions of Datasets and Evaluators }
BikeBench makes numerous key assumptions that may cause significant inaccuracies in design evaluation. A few key assumptions and limitations are listed below, but this is not a comprehensive list. We encourage practitioners using models trained on BikeBench's data or using BikeBench's design evaluation tools for real-world bicycle design to undertake additional validation and verification steps. 
\begin{itemize}
    \item Design representation: BikeBench's design representation is expressive enough to cover a wide variety of designs adhering to a `conventional' diamond-frame bike. It does not span entirely different topological layouts of bicycles, such as recumbent bikes, nor does it model complex suspension systems, such as mountain bike rear suspensions. 
    \item Dataset bias: The original dataset of 4500 human-design bikes features certain biases, particularly in less visually-prominent features like tube thickness values. We refer readers to BIKED~\cite{regenwetter2022biked} and FRAMED~\cite{regenwetter2023framed} for more in-depth discussion. 
    \item Surrogate model robustness: To keep BikeBench fast and portable, we resorted to surrogate models for many of the evaluations. These may lack robustness outside of the support of their data. 
    \item Geometric Feasibility: Our set of closed-form geometric checks is not comprehensive, and primarily covers commonly-seen `mistakes' made by generative models. It is not a guarantee of feasibility. 
    \item Structural evaluation: BIKED includes non-isotropic materials (bamboo, carbon, other) which are difficult to simulate without extensive assumptions. We substitute these materials with aluminum (or steel for carbon fiber) in BikeBench. Real metal tubes nonetheless display some anisotropic properties, causing inaccuracies. Simulations are based on a simplified 3D frame model which approximates tube joints and does not incorporate certain frame features, such as wheel cutouts. Detailed discussion and comparisons to experimental validation are included in FRAMED~\cite{regenwetter2023framed}.
    \item Aerodynamics: The assessment of aerodynamic drag in a constant direct headwind and based on a basic cyclist model is a simplified assessment of bicycle aerodynamics. Real cycling undergoes a variety of different wind speeds at different angles, causing aerodynamic interactions in different flow regimes. 
    \item Aesthetics: Although the trained parametric-to-CLIP embedding model suffices to capture subjective details (e.g. `jet-black triathlon bike')~\cite{regenwetter2025multi}, it struggles to capture technical details of bicycle components (e.g. number of chainrings or fork style). 

\end{itemize}

While these inaccuracies may limit BikeBench's robustness and utility for users interested in physically constructing the bicycles they design, their impact on BikeBench's primary objective---benchmarking---should be minor. 

\subsubsection{Visualization of geometrically infeasible bikes}
\begin{figure}[!htb]
    \centering
    \includegraphics[width=\linewidth]{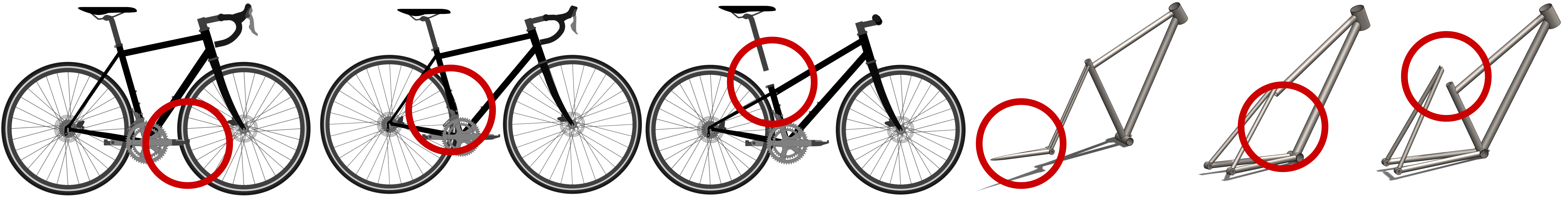}
    \caption{BikeBench features numerous closed-form constraint checks to identify common geometric infeasibilities such as disconnected or colliding components (left 3). Less common infeasibilities in frame geometry are flagged if the design causes errors during 3D reconstruction (right 3).}
    \label{fig:invalidity}
\end{figure}

To give readers some intuition regarding the nature of geometric infeasibility, a few geometrically infeasible bikes are visualized in Figure~\ref{fig:invalidity}.

\subsection{Models and metrics} \label{appx:modelsandmetrics}

\subsubsection{Additional Metrics and Metric Implementation Details}\label{appx:metrics}
\paragraph{Details of Hypervolume Calculation:} In multi-objective design optimization, designs are often compared in the objective space---the space comprising the set of all objective scores. One design ``dominates'' the other if it is superior in every single objective. To calculate the hypervolume metric of a sample set, we measure the hypervolume of the region in the objective space where a corresponding design would be dominated by any design in the sample set. To keep hypervolume bounded, it is typically calculated with respect to some reference point, which we select to be the combination of maximum (worst) objective scores in every unique objective with a small additional margin. Scores are calculated by evaluating every design in BIKED with a random conditional configuration. 

\paragraph{Novelty:} BikeBench also evaluates mean design novelty. We estimate the novelty of any given design as the Euclidean distance to nearest test-set design after scaling designs by test-set per-parameter variance. 

\paragraph{Binary Validity:} While examining overall constraint violation is insightful, designs are ultimately invalid if they violate even a single constraint. Thus, we also evaluate the fraction of designs that simultaneously satisfy all design constraints. 

The public leaderboard will feature these extended scores, in addition to the primary four metrics discussed in the main paper. 

\subsubsection{Details on optimization-augmented model training} \label{appx:oagm_training}
\begin{figure}[!htb]
    \centering
    \includegraphics[width=0.7\linewidth]{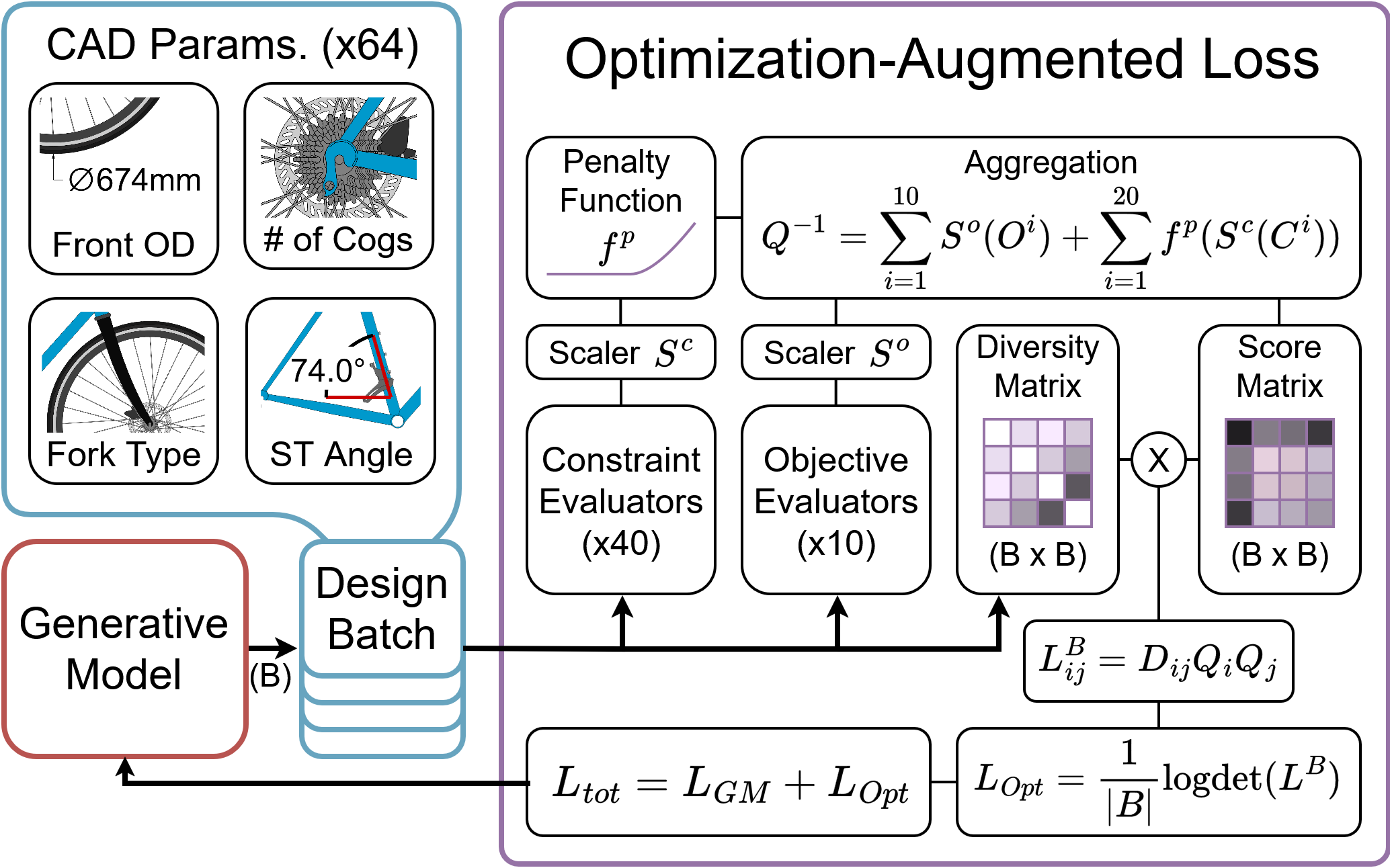}
    \caption{Overview of the optimization-augmented generative model training process.}
    \label{fig:oagm}
\end{figure}

This section presents a detailed discussion of the optimization-augmented generative model training process, with a visual overview in Figure~\ref{fig:oagm}.

The DPP-based auxiliary loss proposed in PaDGAN~\cite{chen2021padgan} uses a single aggregate design quality metric. To condense BikeBench's many objectives and constraints into a differentiable aggregate quality metric, we propose the following aggregation scheme:
\begin{equation}
    s(x) = \sum_{i=1}^{n_o}\frac{o_i(x)}{w_{o_i}} + \sum_{i=1}^{n_c}g(\frac{c_i(x)}{w_{c_i}})
\end{equation}
All $n_o$ objectives, $o_i$, and $n_c$ constraints, $c_i$, are scaled by weighting parameters, $w_{o_i}$ and $w_{c_i}$. Each parameter is set automatically based on the mean absolute value of the scores observed from random pairings of the $n_D$ points in the dataset, $D$, with randomly sampled conditions, $C_j$: 
\begin{equation}
    w_{o_i} = \frac{\sum_{i=0}^{n_D}(|o_i(D_j, C_j)|)}{n_D},\,\, w_{c_i} = \frac{\sum_{i=0}^{n_D}(|c_i(D_j, C_j)|)}{n_D}
\end{equation}
Constraints are additionally fed through a nonlinear scaling function to push them across the constraint boundary and a safe margin away, without rewarding extreme constraint satisfaction:
\begin{equation}
    g(x) = \begin{cases} 
      \frac{\alpha e^{\beta x}}{\beta} & x\leq 0 \\
      \alpha (x + \frac{1}{\beta}) & x \geq 0
   \end{cases}
\end{equation}
This continuous penalty function linearly increases for $x>0$, and morphs into an increasingly gradual decay for $x<0$, and also features a continuous first derivative. We use $\alpha=100$ and $\beta=10$ in our benchmarking. Modulating these terms can encourage a model to focus more on objectives or constraints. For details on the DPP-based loss, given an aggregate quality function, we refer readers to~\cite{chen2021padgan}. 

In the unconditional benchmarks, a small portion of the evaluation budget is held out during training to perform a final rejection sampling pass on a few hundred designs. This (generally) ensures a perfectly valid set. Due to relatively high raw validity rates, relatively few designs have to be passed through the rejection sampling step, making its cost very small (under 10\% of the total evaluation budget). 

The guided DDPM is only applied to the unconditional case, as it directly calls the evaluators on the test-set conditions. The evaluation budget is not high enough to invoke evaluator-based guidance at every denoising step. Instead, the guidance is invoked in a scheduled manner at only certain time steps. 

We refer users interested in detailed training parameters for the OA-GMs to the repository linked in the abstract.

\subsubsection{LLM Prompts} \label{appx:LLMdetails}
A series of prompts is given to the LLM model. These prompts are listed below, excepting a few of the longer subcomponents which are substantiated later in the appendix.
\begin{prompt}[title={Full-Stack LLM Prompts:}]
\begin{itemize}
    \item (System Prompt) ``You are a large language model who will be assisting with a computational bicycle design task. The user will give you some examples of bicycle designs then ask you to create new ones. This is a challenging design task because the designs are represented using csv tables. You'll have to keep good track of the order of the bicycle parameters to make sure you dont forget any or mix any up. The user will also give you some performance metrics for the existing bicycle designs. You will have to use your knowledge of physics and design principles to try to ensure that the designs you generate are valid. Importantly, the user does not want any sort of feedback. They only want the raw final designs in csv format. You'll task is to figure out how to design them to try to meet the design constraints and give them cleanly to the user in csv format. ''
    \item ``I will ask you to create some bicycle designs. The bicycle designs are subject to a set of conditions: a text prompt, some rider dimensions, and a use case. Each design is defined by 64 variables, which I will describe. Some of these are categorical variables, and I will provide you with the possible values for these variables. Others are continuous or boolean. I will describe the design variables shortly. Designs are evaluated according to a set of 50 criteria. 10 of these are objectives and 40 are constraints. Here are the descriptions of the design variables and the evaluation criteria:''
    \item \textit{Dataset descriptions} (see Sec.~\ref{appx:parlist}).
    \item \textit{Evaluation criterion descriptions} (see Sec.~\ref{appx:objlist}).

    \item ``Before asking you to design the bikes, I will provide a dataset of existing bicycle designs to reference. These are useful as a reference point, because it may be difficult to satisfy constraints and objectives if you deviate too far from the space of existing designs. I will also provide a set of objective scores for these bicycle designs. Here are the 25 conditional inputs for the bicycle generation task:''
    \item \textit{Conditional inputs} (see Sec.~\ref{appx:conds}). 
    \item ``Here is a csv file with 25 examples of bikes that satisfy the constraints. The csv has 25 rows (bikes)  plus a header row and exactly 64 columns (parameters):''
    \item \textit{CSV file of valid designs}
    \item ``Next, here are the corresponding scores for those 25 bikes in a csv file of 25 designs (rows)  plus a header row by 50 scores (columns). You can see that all constraint scores are less or equal to zero:''
    \item \textit{CSV file of valid designs' scores}
    \item ``Now that you have seen some example designs that satisfy the constraints, as well as their scores, here is a csv file with 25 examples of bikes that violate the constraints. This csv also has 25 rows (bikes) plus a header row and exactly 64 columns (parameters):''
    \item \textit{CSV file of invalid designs}
    \item ``Finally, here are the corresponding scores for those 25 invalid bikes in a csv file of 25 designs (rows) plus a header row by 50 scores (columns). You can see that at least one constraint score is positive for each:''
    \item \textit{CSV file of invalid designs' scores}

    \item ``Having examined 25 valid and 25 invalid designs and the evaluation of these designs, I hope you have gained an understanding of the design space and design objectives. Please deliberate on a strategy for creating high-performing designs that satisfy the constraints and objectives. Your goal is to create designs that will minimize the objective scores while ensuring every constraint score is less or equal to zero. Here is the prompt for which you will be generating designs:''
    \item \textit{String with the test-set condition}
    \item ``You must create 100 unique bicycle design that satisfy the constraints and objectives. Important: You are not allowed to generate the same bike 100 times. Each design must be unique! You are also encouraged to improve over the valid designs given to you, which may not be very optimal. Please provide the designs in a 100 x 64 csv file with one extra column for the index and one extra row for the column hearers. Please count up from 1-100 in the indices. For the column headers, we recommend you use the parameter names described to help you keep track of what parameter is what. Remember! Every design must be unique. Your final output must be only the csv itself, nothing more. Generate the csv file now.''

\end{itemize}
\end{prompt}

\subsection{Summary of Constraints and Objectives} \label{appx:objlist}
This section summarizes BikeBench's 50 unique constraints and objectives. An overview is presented in Table~\ref{tab:evaluators-summary}, which includes the types of evaluators and the inputs to the evaluation functions. 

\begin{table}[!htb]
\centering
\caption{Summary of BikeBench's evaluation functions, classified as objectives or constraints. Inputs and evaluator type are also specified. }
\label{tab:evaluators-summary}
\resizebox{\textwidth}{!}{%
\begin{tabular}{lcccc}
\toprule
Category     & Objectives & Constraints & Evaluator & Inputs \\
                         \midrule
Geometric Feasibility    & 0          & 32    & Closed-Form + Predictor         & Bike           \\
Structural Soundness     & 4     & 2           & Predictor  & Bike            \\
Aerodynamics             & 1          & 0           & Predictor        & Bike + Rider    \\
Ergonomics               & 3     & 6           & Closed-Form         & Bike + Rider + Use case   \\
Human-Centered Usability & 1     & 0      & Predictor        & Bike           \\
Aesthetics               & 1     & 0           & Predictor     & Bike + Text/Image/Embedding \\
\bottomrule
\end{tabular}%
}
\end{table}

The following is the description of the constraints and objectives given to the LLM, which is also a helpful human-interpretable reference:
\begin{prompt}[title={Constraint and Objective Description Prompt:}]
``Descriptions of all objectives and constraints in the standard BikeBench multi-objective engineering design benchmark. 
Some evaluation criteria are contingent on only the bike design, while others are also contingent on the conditional information. By convention, objectives are minimized at 0, with lower values being better. 
Constraints are also minimized, with 0 being the critical value. Larger positive values indicate more extreme constraint violation and larger magnitude negative values are more constraint-satisfying.
In general, once a constraint is satisfied, we no longer care about further minimizing the value. However, achieving constraints is critical because designs that violate constraints are invalid. Thus, we must attempt to ensure that all constraints are simultaneously below zero. 

The 50 objectives and constraints are described as follows:

\begin{itemize}

\item  Usability Score [Objective]: The predicted 'usability,' as rated by a human, with 0 being the most usable and 1 being the least usable. Predicted by a regression model trained on human-collected ratings.
\item  Drag Force (N) [Objective]: The predicted drag force in N incurred by the cyclist in a 10 m/s headwind, as predicted by a regression model trained on computational fluid dynamics simulation data.
\item  Knee Angle Error (deg.) [Objective]: The difference between the minimum knee angle of the cyclist and the optimal reference range. May include a penalty term if the rider's geometry is completely incompatible with the bike. 
\item  Hip Angle Error (deg.) [Objective]: The difference between the torso-to-upper-leg angle of the cyclist and the optimal reference range. May include a penalty term if the rider's geometry is completely incompatible with the bike.
\item  Arm Angle Error (deg.) [Objective]: The difference between the torso-to-arm angle of the cyclist and the optimal reference range. May include a penalty term if the rider's geometry is completely incompatible with the bike.
\item  Arm Too Long for Bike [Constraint]: Constraint indicating that the length of the rider's arm is longer than the length of the rider's torso plus the distance from the saddle to the handlebars. Means the saddle is too close to the handlebars. 
\item  Saddle Too Far From Handle [Constraint]: Constraint indicating that the distance from the saddle to the handlebars is longer than the length of the rider's torso plus the length of the rider's arm. Means the saddle is too far from the handlebars. 
\item  Torso Too Long for Bike [Constraint]: Constraint indicating that the length of the rider's torso is longer than the length of the rider's arm plus the distance from the saddle to the handlebars. Means the saddle is too close to the handlebars. 
\item  Saddle Too Far From Crank [Constraint]: Constraint indicating that the distance from the saddle to the pedals in the far position is longer than the length of the rider's upper leg plus the length of the rider's lower leg. Means the saddle is too far from the pedals. 
\item  Upper Leg Too Long for Bike [Constraint]: Constraint indicating that the length of the rider's upper leg is longer than the distance from the saddle to the pedals in the far position plus the length of the rider's lower leg. Means the saddle is too close to the pedals. 
\item  Lower Leg Too Long for Bike [Constraint]: Constraint indicating that the length of the rider's lower leg is longer than the distance from the saddle to the pedals in the far position plus the length of the rider's upper leg. Means the saddle is too close to the pedals. 
\item  Cosine Distance To Embedding [Objective]: The cosine distance in the CLIP embedding space between the rendered bike image and the target text or image embedding.
\item  Mass (kg) [Objective]: The mass in kg of the bike frame, as predicted by a regression model trained on finite element analysis data.
\item  Planar Compliance Score [Objective]: A composite planar compliance score for the bike frame, as predicted by a regression model trained on finite element analysis data.
\item  Transverse Compliance Score [Objective]: A composite transverse compliance score for the bike frame, as predicted by a regression model trained on finite element analysis data.
\item  Eccentric Compliance Score [Objective]: A composite eccentric compliance score for the bike frame, as predicted by a regression model trained on finite element analysis data.
\item  Planar Safety Factor [Constraint]: Constraint quantified as 1.5 minus the safety factor under planar loading, as predicted by a regression model trained on finite element analysis data. Means the frame fails under planar loading.
\item  Eccentric Safety Factor [Constraint]: Constraint quantified as 1.5 minus the safety factor under eccentric loading, as predicted by a regression model trained on finite element analysis data. Means the frame fails under eccentric loading.
\item  Predicted Frame Validity [Constraint]: Constraint indicating some abstract issue with the frame, as predicted by a classification model trained to identify CAD models that failed to regenerate. Means the frame is invalid in an unspecified way.
\item  Saddle Height Too Small [Constraint]: Constraint indicating that the saddle height is too low to be functional.
\item  Saddle Collides With Seat Tube [Constraint]: Constraint indicating that the saddle height is so low that it collides with the top of the seat tube.
\item  Saddle Too Short [Constraint]: Constraint indicating that the length of the saddle is too short. 
\item  Head Angle Over Limit [Constraint]: Constraint indicating that the head angle is over 180 degrees. 
\item  Seat Angle Over Limit [Constraint]: Constraint indicating that the seat angle is over 180 degrees. 
\item  Seat Post Too Short [Constraint]: Constraint indicating that the seat post doesn't reach the seat tube given the prescribed saddle height.
\item  Seat Post Too Long [Constraint]: Constraint indicating that the seat post is so long that it hits the bottom bracket given the prescribed saddle height.
\item  Rear Wheel Inner Diameter Too Small [Constraint]: Constraint indicating that the rear wheel's inner diameter is too small to be functional.
\item  Front Wheel Inner Diameter Too Small [Constraint]: Constraint indicating that the front wheel's inner diameter is too small to be functional.
\item  Seat Tube Extension Longer Than Seat Tube [Constraint]: Constraint indicating that the distance from the top of the seat tube to the top tube junction is longer than the seat tube itself.
\item  Head Tube Upper Extension And Lower Extension Overlap [Constraint]: Constraint indicating that the top tube and down tubes intersect before their junctions with the head tube.
\item  Seat Stay Junction Longer Than Seat Tube [Constraint]: Constraint indicating that the distance from the top of the seat tube to the seat stay junction is longer than the seat tube itself.
\item  Non-negative Parameter Is Negative [Constraint]: Constraint indicating that at least one parameter that should be strictly positive is negative.
\item  Chain Stay Smaller Than Rear Wheel Radius [Constraint]: Constraint indicating that the chain stay length is smaller than the wheel radius, creating a collision.
\item  Chain Stay Shorter Than BB Drop [Constraint]: Constraint indicating that the vertical drop from the rear axle to the bottom bracket is greater than the chain stay length, creating an impossibility. 
\item  Seat Stay Smaller Than Rear Wheel Radius [Constraint]: Constraint indicating that the seat stay length is smaller than the wheel radius, creating a collision.
\item  Seat Tube Collides With Rear Wheel [Constraint]: Constraint indicating that the seat tube collides with the rear wheel. 
\item  Down Tube Can't Reach Head Tube [Constraint]: Constraint indicating that the down tube isn't long enough to reach the head tube, creating an impossibility. 
\item  Rear Wheel Cutout Severs Seat Tube [Constraint]: Constraint indicating that the diameter of the aero tube cutout of the seat tube is so large such that it is completely severing the seat tube. 
\item  Foot Collides With Front Wheel [Constraint]: Constraint indicating that the foot would collide with the pedal in its forward position, causing a collision when turning. 
\item  Crank Hits Ground In Lowest Position [Constraint]: Constraint indicating that the crank hits the ground during its rotation.
\item  RGB Value Greater Than 255 [Constraint]: Constraint indicating that one or more frame RGB values exceed 255.
\item  Chain Stays Collide [Constraint]: Constraint indicating that the chain stays collide with each other before reaching the bottom bracket.
\item  Tube Wall Thickness Exceeds Radius [Constraint]: Constraint indicating that some tubes' wall thickness exceeds their radius, creating an impossibility.
\item  Seat Tube Too Narrow For Seat Post [Constraint]: Constraint indicating that the seat post's outer diameter is wider than the seat tube's inner diameter, creating a collision. 
\item  Down Tube Improperly Joins Head Tube [Constraint]: Constraint indicating that the down tube is partially disconnected from the bottom of the head tube.
\item  Top Tube Improperly Joins Head Tube [Constraint]: Constraint indicating that the top tube is partially disconnected from the top of the head tube.
\item  Top Tube Improperly Joins Seat Tube [Constraint]: Constraint indicating that the top tube is partially disconnected from the top of the seat tube.
\item  Down Tube Collides With Front Wheel [Constraint]: Constraint indicating that the down tube collides with the front wheel. 
\item  Saddle Hits Top Tube [Constraint]: Constraint indicating that the saddle collides with the top tube.
\item  Saddle Hits Head Tube [Constraint]: Constraint indicating that the saddle collides with the head tube.''
    
\end{itemize}
\end{prompt}
\subsection{Summary of Design Variables} \label{appx:parlist}

This section contains details on BikeBench's parametric design representation. The following is the description of the design variables given to the LLM, which is also a helpful human-interpretable reference:

\begin{prompt}[title={Dataset Description Prompt:}]
``Descriptions of all parameters in the standard BikeBench design representation scheme. 
General notes: All lengths are measured in mm. All angles are measured in degrees.

The 64 variables are described as follows:
\begin{enumerate}
\item  'Seatpost LENGTH' [Continuous]: The length of the seat post.
\item  'CS textfield' [Continuous]: The length of the chain stay tubes.
\item  'BB textfield' [Continuous]: Bottom bracket drop, measured as the vertical drop from the rear axle to the center of the bottom bracket. By convention, positive values imply the bottom bracket lies below the axle.
\item  'Stack' [Continuous]: The vertical distance from the top of the head tube to the bottom bracket.
\item  'Head angle' [Continuous]: The angle of the head tube clockwise from horizontal, in degrees.
\item  'Head tube length textfield' [Continuous]: The length of the head tube.
\item  'Seat stay junction0' [Continuous]: The length along the seat tube from the top of the seat tube to the junction with the seat stays. By convention, this is measured to the center of the seat stays.
\item  'Seat tube length' [Continuous]: The length of the seat tube.
\item  'Seat angle' [Continuous]: The angle of the seat tube clockwise from horizontal.
\item  'DT Length' [Continuous]: The length of the down tube.
\item  'FORK0R' [Continuous]: Fork offset, measured as the perpendicular distance from the front axle to the head tube axis.
\item  'BB diameter' [Continuous]: The diameter of the bottom bracket.
\item  'ttd' [Continuous]: Top tube outer diameter.
\item  'dtd' [Continuous]: Down tube outer diameter.
\item  'csd' [Continuous]: Chain stay outer diameter.
\item  'std' [Continuous]: Seat tube outer diameter.
\item  'htd' [Continuous]: Head tube outer diameter.
\item  'ssd' [Continuous]: Seat stay outer diameter.
\item  'Chain stay position on BB' [Continuous]: The distance along the length of the bottom bracket from its edge to the center of the chain stay tubes.
\item  'SSTopZOFFSET' [Continuous]: The offset from the center plane of the bike of the joints connecting the seat stays to the seat tube.
\item  'MATERIAL' [Categorical]: The material of the bike frame. Possible values are: 'ALUMINIUM', 'STEEL', 'TITANIUM'.
\item  'Head tube upper extension2' [Continuous]: The length from the top of the head tube to the junction with the top tube. By convention, this is measured to the center of the top tube.
\item  'Seat tube extension2' [Continuous]: The length from the top of the seat tube to the junction with the top tube. By convention, this is measured to the center of the top tube.
\item  'Head tube lower extension2' [Continuous]: The length from the bottom of the head tube to the junction with the down tube. By convention, this is measured to the center of the down tube.
\item  'SEATSTAYbrdgshift' [Continuous]: The distance along the center plane of the bike from the seat stay and seat tube junction to the seat stay bridge, if present on the bike.
\item  'CHAINSTAYbrdgshift' [Continuous]: The distance along the center plane of the bike from the outer edge of the bottom bracket to the chain stay bridge, if present on the bike.
\item  'SEATSTAYbrdgdia1' [Continuous]: The diameter of the seat stay bridge, if present on the bike.
\item  'CHAINSTAYbrdgdia1' [Continuous]: The diameter of the chain stay bridge, if present on the bike.
\item  'SEATSTAYbrdgCheck' [Boolean]: A boolean value indicating whether the seat stay bridge is present on the bike.
\item  'CHAINSTAYbrdgCheck' [Boolean]: A boolean value indicating whether the chain stay bridge is present on the bike.
\item  'Dropout spacing' [Continuous]: The distance between the rear dropouts.
\item  'Wall thickness Bottom Bracket' [Continuous]: The tube wall thickness of the bottom bracket.
\item  'Wall thickness Top tube' [Continuous]: The tube wall thickness of the top tube.
\item  'Wall thickness Head tube' [Continuous]: The tube wall thickness of the head tube.
\item  'Wall thickness Down tube' [Continuous]: The tube wall thickness of the down tube.
\item  'Wall thickness Chain stay' [Continuous]: The tube wall thickness of the chain stay.
\item  'Wall thickness Seat stay' [Continuous]: The tube wall thickness of the seat stay.
\item  'Wall thickness Seat tube' [Continuous]: The tube wall thickness of the seat tube.
\item  'Wheel diameter front' [Continuous]: The outer diameter of the front wheel.
\item  'RDBSD' [Continuous]: The difference between rear wheel outer diameter and bead seat diameter, roughly approximating the tire thickness.
\item  'Wheel diameter rear' [Continuous]: The outer diameter of the rear wheel.
\item  'FDBSD' [Continuous]: The difference between front wheel outer diameter and bead seat diameter, roughly approximating the tire thickness.
\item  'Fork type' [Categorical]: The style of fork. Possible values are: '0', '1', '2'. 0 is a rigid fork, 1 is a single crown fork, 2 is a double crown fork.
\item  'Stem kind' [Categorical]: The style of stem. Possible values are: '0', '1', '2'. 0 is a stem that features a sharp and immediate angle away from the head tube. 1 is a stem that features a sharp angle some distance away from the head tube. 2 is a stem that features a gradual angle away from the head tube after initially extending in line with the head tube.
\item  'Handlebar style' [Categorical]: The style of the handlebars. Possible values are: '0', '1', '2'. 0 is a drop bar, 1 is a mountain bike bar, 2 is a bullhorn bar.
\item  'BB length' [Continuous]: The length of the bottom bracket.
\item  'Wheel cut' [Continuous]: The diameter of the cutout of the seat tube for the rear wheel, if using an aerodynamic tube type.
\item  'BELTorCHAIN' [Boolean]: A boolean value indicating whether the bike has a chain (True) as opposed to a belt.
\item  'Number of cogs' [Integer]: The number of cogs on the rear wheel.
\item  'Number of chainrings' [Integer]: The number of chainrings attached to the crank.
\item  'FIRST color R\_RGB' [Continuous]: The red component of the primary paint color of the bike.
\item  'FIRST color G\_RGB' [Continuous]: The green component of the primary paint color of the bike.
\item  'FIRST color B\_RGB' [Continuous]: The blue component of the primary paint color of the bike.
\item  'RIM\_STYLE front' [Categorical]: The style of the front rim. Possible values are: 'DISC', 'SPOKED', 'TRISPOKE'. Despite the name, trispoke class implies composite spokes but does not necessarily imply three composite spokes.
\item  'RIM\_STYLE rear' [Categorical]: The style of the rear rim. Possible values are: 'DISC', 'SPOKED', 'TRISPOKE'. Despite the name, trispoke class implies composite spokes but does not necessarily imply three composite spokes.
\item  'SPOKES composite front' [Integer]: If applicable, the number of composite spokes in the front wheel minus two (a value of 1 is a trispoke wheel).
\item  'SPOKES composite rear' [Integer]: If applicable, the number of composite spokes in the rear wheel minus two (a value of 1 is a trispoke wheel).
\item  'SBLADEW front' [Continuous]: If applicable, the width of the front wheel composite spokes.
\item  'SBLADEW rear' [Continuous]: If applicable, the width of the rear wheel composite spokes.
\item  'Saddle length' [Continuous]: The length of the saddle.
\item  'Saddle height' [Continuous]: The vertical distance from the saddle to the bottom bracket.
\item  'Seat tube type' [Categorical]: The style of seat tube. Possible values are: '0', '1', '2'. 0 is aerodynamic, while 1 and 2 are standard round tubes with no distinction in this representation scheme.
\item  'Head tube type' [Categorical]: The style of head tube. Possible values are: '0', '1', '2', '3'. 0 is aerodynamic, while 1 and 2 are standard round tubes with no distinction in this representation scheme. 3 is a tapered head tube.
\item 'Down tube type' [Categorical]: The style of down tube. Possible values are: '0', '1', '2'. 0 is aerodynamic, while 1 and 2 are standard round tubes with no distinction in this representation scheme. ''
\end{enumerate}
\end{prompt}

\subsection{Conditional Tasks}\label{appx:conds}

Conditional information is concatenated for the LLM into a text string of the form: `Rider Body Dimensions: Upper leg length - [ULL], Lower leg length - [LLL], Arm length - [AL], Torso length - [TL], Neck and head length - [HNL], Torso width - [TW]. Use Case: [Road Biking/Mountain Biking/Commuting]. Bike Description: [Text Prompt]'.

An example condition from the test set is included below: 
\begin{prompt}[title={Example Test Set Condition in Text String Representation:}]
``Rider Body Dimensions: Upper leg length - 0.3772640228, Lower leg length - 0.5058981180, Arm length - 0.6180832386, Torso length - 0.4945643246, Neck and head length - 0.3314186633, Torso width - 0.3427633643. Use Case: Road Biking. Bike Description: Track training bike with fixed gear, deep-section rims, and a stiff frame designed to transfer power effectively to the track.''
\end{prompt}

\subsection{Computing Resources and Cost}
All experiments were carried out on a workstation with an RTX 3090Ti GPU and a Ryzen 5900x CPU. No model training or optimization took more than 4 hours. API calls for GPT-5 queries were divided over the 100 test conditions and parallelized across 20 processes. Each LLM benchmark took approximately 1 hour to complete and cost approximately 40 USD. 

\subsection{Ethics and Societal Impact}
BikeBench aims to advance the capabilities of generative AI models for engineering design. In general, AI has many positives and negatives, most of which are not particularly pertinent to this work. However, we would like to acknowledge some of the pros and cons of generative models for engineering design. A principal risk of AI in engineering, particularly generative models, is safety. Generative models are usually probabilistic. When human safety is in question, even small chances of failure are unacceptable. Designs and design decisions created by AI must be held to the same (or higher) engineering standards as human designs. Engineering design AI also stands poised to deliver notable societal benefits. By lowering the barrier of entry to design, it may democratize design, allowing individuals without formal training to participate in the design process. By increasing design throughput, it may realize better products, more efficient design, and greater design customization for individuals. We advocate for the careful and ethical use of AI in engineering design and beyond.

\end{document}